\documentclass[fleqn,usenatbib,useAMS]{mnras}

\usepackage{mathptmx}

\usepackage[T1]{fontenc}
\usepackage{ae,aecompl}

\usepackage{graphicx}	
\usepackage{amsmath}	
\usepackage{amssymb}	
\usepackage{multicol}        
\usepackage{bm}		
\usepackage{pdflscape}	
\usepackage{appendix}

\usepackage{color}
\usepackage{soul}
\usepackage{txfonts}
\usepackage{dcolumn}

\newcommand{\arcdeg}{$^{\circ}$}

\usepackage[export]{adjustbox}
\newcolumntype{d}{D{.}{.}{-1}}

\usepackage{lastpage}

\title[Identification of asteroids using the VO]{Identification of asteroids using the Virtual Observatory: the WFCAM Transit Survey}

\author[Cort\'es-Contreras et al.]{
M. Cort\'es-Contreras,$^{1,2}$\thanks{E-mail: mcortes@cab.inta-csic.es}
F. M. Jim\'enez-Esteban$^{1,2}$,
M. Mahlke$^{1,2}$,
E. Solano$^{1,2}$,
\newauthor
J. \v{D}urech$^{3}$,
S. Barcel\'o Forteza$^{1}$,
C. Rodrigo$^{1,2}$,
A. Velasco$^{1,2}$
and B. Carry$^{4}$
\\
$^{1}$Departmento de Astrof\'{\i}sica, Centro de Astrobiolog\'{\i}a (CSIC-INTA), ESAC Campus, Camino Bajo del Castillo s/n, E-28692 Villanueva de la Ca\~nada, Madrid, Spain\\
$^{2}$Spanish Virtual Observatory, Spain\\
$^{3}$Astronomical Institute, Faculty of Mathematics and Physics, Charles University, V Hole\v{s}ovi\v{c}k\'ach 2, 180\,00 Prague 8, Czech Republic\\
$^{4}$Universit\'e C\^ote d'Azur, Observatoire de la C\^ote d'Azur, CNRS, Laboratoire Lagrange, France
}

\date{Accepted XXX. Received YYY; in original form ZZZ}

\pubyear{2019}

\begin{document}
\label{firstpage}
\pagerange{\pageref{firstpage}--\pageref{lastpage}}
\maketitle

\begin{abstract}

{The nature and physical properties of asteroids, in particular those orbiting in the near-Earth space, are of scientific interest and practical importance.
Exoplanet surveys can be excellent resources to detect asteroids, both already known and new objects. This is due their similar observing requirements: large fields of view, long sequences, and short cadence.
If the targeted fields are not located far from the ecliptic, many asteroids will cross occasionally the field of view.
We present two complementary methodologies to identify asteroids serendipitously observed in large-area astronomical surveys. 
One methodology focuses on detecting already known asteroids using the Virtual Observatory tool SkyBoT, which predicts their positions and motions in the sky at a specific epoch. The other methodology applies the \textit{ssos} pipeline, which is able to identify known and new asteroids based on their apparent motion. The application of these methods to the 6.4 $deg^{2}$ of the sky covered by the Wide-Field CAMera Transit Survey in the J-band is described. We identified 15\,661 positions of 1\,821 different asteroids. Of them, 182 are potential new discoveries. A publicly accessible online, Virtual Observatory compliant catalogue was created. We obtained the shapes and periods for five of our asteroids from their light-curves built with additional photometry taken from external archives. We demonstrated that our methodologies are robust and reliable approaches to find, at zero cost of observing time, asteroids observed by chance in astronomical surveys. Our future goal is to apply them to other surveys with adequate temporal coverage.}

\end{abstract}




%

\section{Introduction}
Small Solar System bodies were defined in 2006 by the IAU as those objects that are neither planets nor dwarf planets, nor satellites of a planet or a dwarf planet. As of July 2019, more than 790\,000 small Solar System bodies are known\footnote{\url{https://www.minorplanetcenter.net/}}. The large majority of them are asteroids. They occupy a variety of orbits ranging from near-Earth to the Kuiper Belt. Their study is motivated, among other reasons, by their intrinsic importance as remnants of the early stages of the solar system formation process \citep{demeo}, as well as by practical reasons concerning space exploration or the impact frequency with Earth \citep[e.g.,][]{Ches02, Spoto18}.

An accurate determination of the asteroid orbital parameters is crucial to assert a reliable probability of a future collision with our planet. This is only possible if good measurements of the sky position of the asteroid exist at different epochs and these measurements cover a large fraction of the asteroid orbit. In this sense, searching for fortuitous detection of asteroids in archive images can help in extending the part of the orbit covered by the observations \citep{Boattini01, Gwyn12, Solano14}.

Together with accurate orbital parameters, other physical parameters are important to properly characterize an asteroid. From the analysis of the changes on the asteroid's brightness due to changes in its geometry, one can derive the rotational period, the scattering properties of the surface, whether the asteroid is spinning or not around its major axis, or even binary nature \citep{durech15, margot15}. Another property that can be determined from the analysis of its light-curve is the asteroid's shape. Density is critical to estimate the real threat of a potential collision with the Earth. While masses can be determined using different methodologies (see for instance \citealt{Carry12}), the major uncertainties in density estimations come from the uncertainties in the volume \citep{scheeres15}. Therefore, precise reconstruction of the 3-D shape is important. To all this, the recovery of photometric measurements from archive data of asteroids serendipitously observed can help, especially for those surveys with long sequences and short cadence.

These properties are key in triggering the dynamical evolution of small asteroids through the Yarkovsky effect \citep{Vou15}, which slowly change the semi-major axis of their orbits and put them in resonances with giant planets, injecting them on planet-crossing orbits \citep{Gran18}.

Because of the above described reasons, there has been an increasing trend of exploiting large sets of images for the discovery and characterization of asteroids accidentally observed, especially those located not far from the ecliptic plane. 
For example, \cite{Popescu16} built the {\it MOVIS} catalogue by recovering near 40\,000 SSOs in the near infrared VISTA-VHS survey \citep{Cross12}. Two years later, they increased the sample of detected moving objects in more than 10\,000 and provided a taxonomic classification in \cite{Popescu18}.
A similar work was carried out by \cite{mahlke2018mining}, who recovered  about 20\,000 SSOs in the KiDS \citep{kids2017} optical survey.
Another relevant work in the field is that of \cite{Vaduvescu2017}, who data-mined Suprime Cam \citep{SuprimeCam} images for Near-Earth asteroids (NEAs). They recovered more than 2\,500 asteroids, one fifth of them being NEAs.
In the future, it is expected the ESA {\it Euclid} mission to observe about 150\,000 SSOs at high inclinations (i > 15\arcdeg), which makes it a precious resource for Solar System research \citep{Carry18}. All these projects reflect a rising interest on recovering information harboured in large-area surveys or even in particular sets of images, and the need of elaborating methodologies and developing tools aiming to help their retrieval.

Among wide imaging surveys, exoplanet surveys are excellent resources to get light-curves of asteroids \citep[e.g.,][]{SzaboR2016, Molnar2018}. This is because both types of targets share similar observing requirements: large field of views (FOV), long sequences, and short cadences. For this reason and the accessibility of the data, the Wide-Field CAMera \citep[WFCAM,][]{Casali07} Transit Survey \citep[WTS,][]{wts} was chosen.

This article is organised as follows: in Sect. \ref{sec:WTS}, we describe the WFCAM Transit Survey; Sect. \ref{sec:methodology} presents the approaches that we followed to identify the asteroids and calibrate their light-curves; in Sect. \ref{sec:results}, we discuss the results obtained; and Sect. \ref{sec:conclusions} contains the conclusions of this work.

\section{The WFCAM Transit Survey}
\label{sec:WTS}

The infrared WFCAM is an instrument mounted at the 3.8-m United Kingdom Infrared Telescope (UKIRT). The WTS was awarded 200 nights of observations with UKIRT to perform the first ever systematic near-infrared search for transiting exoplanets around cool dwarfs. Observations started in August 2007 and lasted till March 2013, ending up with over 600 observing nights at the end of the programme. They were carried out when observing conditions did not meet the restrictive criteria of other surveys (e.g. seeing $>$\,1\arcsec). As a consequence, the observations were not uniformly distributed over time. The survey, as well as the reduction process, are described in detail in \cite{Kovacs13}. The exploitation of the WTS was accomplished in the framework of the RoPACS\footnote{\url{http://star.herts.ac.uk/RoPACS/}} 
(Rocky Planets Around Cool Stars) - Marie Curie Initial Training Network.

WTS targeted four fields of 1.6 square degrees each, mainly in J-band.
In this band, observing runs took place during a variable fraction of the night, ranging between a few minutes and almost 12 hours. Accordingly, the number of observations per night varies from 2 to 241, with an average of 51 observations per night. Telescope dedicated time is usually around 1.3 hours per night and the typical cadence in the observations is around 2\,min.
Typical exposure times range from 5 to 10 seconds.
The high quality of the images, the large FOV of WFCAM (four 2048$\times$2048 Rockwell Hawaii-II PACE arrays covering 13.65\arcmin$\times$13.65\arcmin\ each), the high spatial resolution (plate scale of 0.4\arcsec), together with the observing strategy with typically several tens of exposures of the same field in the same night, make WTS a good resource to identify and characterize moving sources. 

\begin{figure}
  \centering
   \includegraphics[width=\hsize,trim={1.9cm 2.6cm 0.9cm 3.3cm},clip]{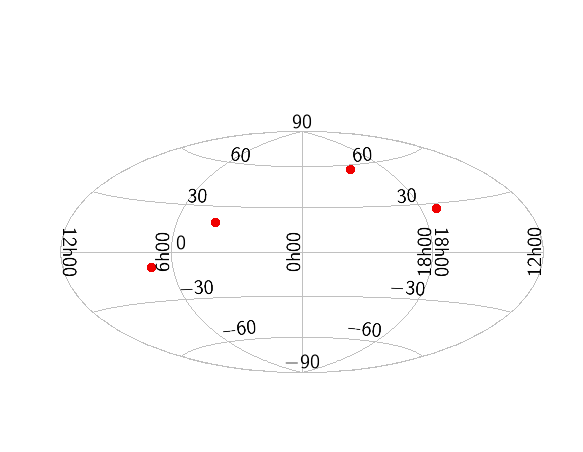}
      \caption{Aitoff projection of the sky position in ecliptic coordinates of the four observed fields by the WTS.}
         \label{fig.sky-map}
\end{figure}

In addition, three out of four of the observed regions by the WTS are below 30\arcdeg\ of the ecliptic plane, where most of the known asteroids are distributed. Figure~\ref{fig.sky-map} shows the position of the observed fields in ecliptic coordinates. Table~\ref{tab.regions} lists the central ecliptic coordinates of each region, the number of observing nights, and the total number of images taken during the programme. The number of asteroids and asteroid counterparts found in each WTS field separated by asteroid dynamical class are also presented in this table (see Sect.\ref{sec:results}).

\begin{table*}
        \centering
        \caption {Ecliptic coordinates of the center of the four regions observed by the WTS, the number of observing nights, and the total number of scientific images taken at each region along the whole survey, together with the number of asteroids and asteroid's  detections grouped by dynamical class and WTS field (see Sect\,\ref{sec:results}).}
        \label{tab.regions}
        \begin{tabular}{c c c c c c c c c c } 
        \hline \hline
        \noalign{\smallskip}
\multicolumn{2}{c}{Ecliptic coordinates (\arcdeg)} & Obs.   & Number     &	\multicolumn{6}{c}{Asteroids found (Asteroid's detections)} \\
Longitude &	Latitude & nights & of images	&	Main Belt	&	Hungaria	&	Mars-Crosser	&	Trojan	&	NEA  & Unknown    \\

        \noalign{\smallskip}
        \hline
        \noalign{\smallskip}
        \noalign{\smallskip}
106.16	& ~~--9.59  & 161 & ~~6\,499 &	1\,098 (10\,440)	&	17 (120)	&	14 (110)	&	34	(405)	&	1 ~~(3) & 67 (265)	\\
~~61.75	& +19.19	& 142 & ~~5\,396 &	~~~322 ~~(2\,489)	&	13 ~~(65)	&	~~5  ~~(52)	&	~~4 ~~(30)	&	5 (15) & 40 (155)	\\
257.41	& +26.69	& 204 & ~~8\,043 &	~~~~~91 ~~~~(978)	&	22 (191)	&	~~6  ~~(34)	&	~~4 ~~(64)	&	2 ~~(6) & 74 (230) 	\\
306.34	& +56.93	& 291 &  10\,939 &	--	&	--	&	~~1 ~~~~(5)	&	--	&	-- & ~~1 ~~~~(4)	\\

\noalign{\smallskip}    
        \hline
        \end{tabular}
\end{table*}


\section{Methodology}
\label{sec:methodology}

Two different and complementary methods were used to identify the asteroids in the WTS images. This allowed us to compare and validate the results obtained with both methods.

\subsection{Sighted method: SkyBoT}

This method searched for detections of already known asteroids in the WST J-band images. It is based on the prior information obtained from the Virtual Observatory (VO) compliant service SkyBoT\footnote{\url{http://vo.imcce.fr/webservices/skybot/}} (Sky Body Tracker, \citealt{Ber06,Ber16}) and uses the following procedure.

\subsubsection{Image cleaning}\label{sec:img_clean}

We gathered 30\,558 bias and flat corrected, and astrometrically calibrated WTS images in the J-band from the WFCAM Science Archive\footnote{\url{http://wsa.roe.ac.uk/}}. We used Aladin \citep{Bonna00} to extract the subimage taken by each of the four detectors and managed them independently in the following steps. In the edges of the detectors, the signal significantly decreases, and the upper left corners are noticeably noisy. In order to minimize the number of false detections during the source extraction process, we removed the pixels of these regions by trimming the subimages in the first 29 and the last 40 pixels in the $x$ direction and in the first 37 and the last 45 pixels in the $y$ direction. This adds a total of 69 and 82 pixels removed in the $x$ and $y$ directions, accordingly. In addition, we discarded every detection at the upper-left corner ($y > 2.1624x$+$1934.34$).

We noticed that for 770 images taken on the 23th of February 2010, on the 8th and 9th of January 2012, and on the 3th and 4th of February 2014, the subimages corresponding to the third detector were blank.
Thus, in total we searched for asteroids in 121\,462 subimages, all in the J band.

\subsubsection{Source catalogues}

We detected sources by running {\sc SExtractor} \citep{Bertin96} on every trimmed subimage. The {\sc SExtractor} configuration parameters used in this analysis are summarized in Table~\ref{table.sexparam} in Appendix~\ref{app.config}. This way we constructed a {\sc SExtractor} catalogue for each subimage. Using these {\sc SExtractor} catalogues and the {\it Gaia} DR2 catalogue as a reference, we estimated an average astrometric error for the WTS images of 0.15\arcsec\ ($\sigma_{WTS}$). 

\begin{figure}
        \centering
        \includegraphics[width=\hsize]{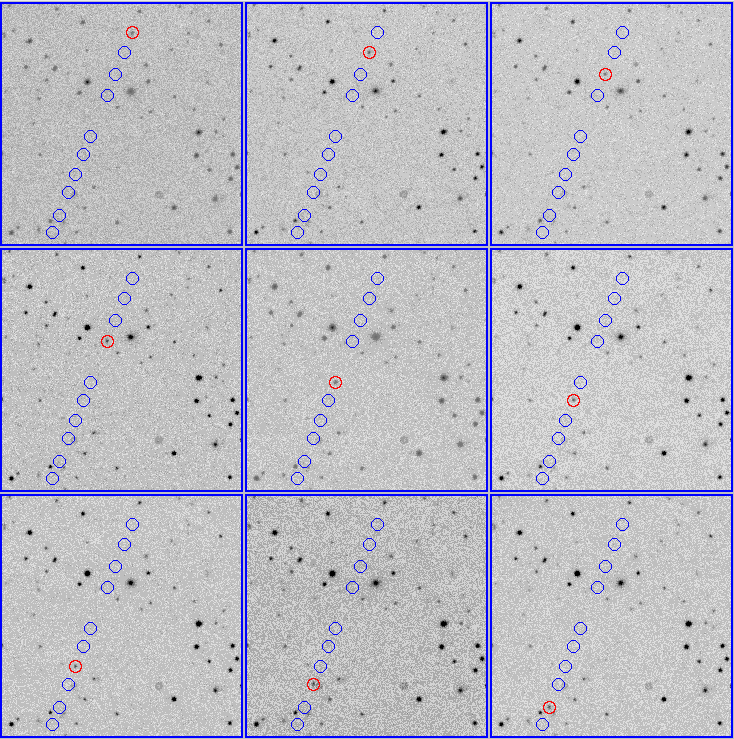}
        \caption{Example of linear motion of the Main Belt asteroid 1997 UK9 during nine consecutive images centered at 17:17:39.5 +04:04:15.1. The temporal coverage is almost 3\,h. The epoch increases from left to right and from top to bottom in intervals of near 20 minutes. The proper motion of the asteroid is 42.8\arcsec/h. Blue circles indicate the trajectory of the asteroid in a $\sim$2.5 square degrees sized field. Red circles indicate the current asteroid position in the image.}
         \label{fig.astmotion}
\end{figure}

In a sequence of images taken during the same night, asteroids appear as moving sources (see an example in Figure~\ref{fig.astmotion}). The typical value for a main-belt object (MBA) is 18\arcsec/h, while NEAs can be as fast as tens of arcseconds per hour or even faster when they come close to Earth. To distinguish between asteroids and any other source in the field, we built a catalogue composed by all {\sc SExtractor} sources detected at the same position in different images within an error of 0.4\arcsec\ ($\sim$\,3\,$\times$\,$\sigma_{WTS}$).
This catalogue covers the four regions observed in the WTS and contains 1\,049\,284 unique entries, mainly celestial sources but also bad pixels and artifacts.
Note that asteroid detections lying within 0.4\arcsec, either because they are slow motion objects or because the time-lapse between images is too short, would be treated as non-moving sources. They would therefore be non-detected asteroids.
From now on, we will refer to this catalogue as the {\it Non-moving Source Catalogue}. 

\subsubsection{Photometric calibration}
\label{phot}

To build the light-curves of our asteroids, we need to photometrically calibrate our images. Because the UKIDSS Large Area Survey, which shares photometric system with the WTS, does not cover any of the WTS observed fields, we used the 2MASS catalogue instead.

We selected all 2MASS counterparts separated by less than 0.5\arcsec\ from the {\sc SExtractor} field sources, with good photometry (i.e., quality flag {\tt A} in the J-band), and being fainter than $J$\,=\,10.5\,mag in order to avoid saturation in the WTS images. We performed an iterative linear fit in which only photometric points deviating by less than 3$\sigma_{cal}$ remain, being $\sigma_{cal}$ the standard deviation of the difference between the J-band calibrated magnitude of the {\sc SExtractor} field sources and their 2MASS counterparts. The number of 2MASS sources used for calibration varies from 135 to 1188 per subimage.

The photometric zero point determines the connection between the observed counts and the 2MASS photometric system in the form:
\begin{equation}
    m_*=m_{SExtractor}+m_{ZP}
\end{equation}
where $m_*$ is the calibrated magnitude, $m_{SExtractor}$ is the instrumental magnitude from {\sc SExtractor} and $m_{ZP}$ is the zero point magnitude. The linear fit carried our for the calibration provides us the zero point. The slope takes values near unity and the zero point magnitude takes a typical value of 25.279\,mag with an average associated error of 0.017\,mag.

Although it was necessary to extrapolate for objects fainter than 16\,mag in J-band (2MASS magnitude limit), the errors associated to the calibration beyond that limit have a typical value of 0.03\,mag. Hence, the greatest contribution to the photometric errors are the uncertainties in the photometry provided by {\sc SExtractor}. Thus, the estimated typical error in the calibrated magnitudes is 0.11\,mag, reaching 0.18\,mag for J\,$\geq $\,19\,mag. We want to highlight here that calibrations between consecutive images within the same night and between consecutive nights are consistent within the errors.

For a determination of the limiting magnitude of the survey, we considered the most frequent calibrated magnitude in every image. On average, the magnitude limit of the images stands at 18.5\,mag.
We also observe that sources under 10.7\,mag saturate.

\subsubsection{Asteroid identification}\label{sec:ast_ind}

We identified the asteroids lying in the FOV of our images at the epoch of observation making use of SkyBoT. This Virtual Observatory (VO) service provides a fast and simple cone-search method to list all known asteroids within a given region of the sky at a given epoch. For that, it weekly precomputes ephemerides based on their osculating elements computed at the Lowell observatory (ASTORB database\footnote{\url{https://asteroid.lowell.edu/main/astorb}}).
We queried SkyBoT with a search radius of 0.17\arcdeg\ centered in each subimage (for comparison, the diagonal of the detector field is 0.32\arcdeg). This radius is enough to entirely cover the whole subimage sky area. We used the default {\tt -filter} parameter, which limits the search to asteroids with position errors under 120\arcsec.
Among other parameters, SkyBoT provides information on the identification name and the dynamical class of the asteroid, as well as its expected position in the sky with the corresponding errors ($\sigma_{SBT}$), the estimated V magnitude, and the proper motion, at the requested epoch.

For each subimage, we obtained a list of candidate asteroids by cross-matching the corresponding {\sc SExtractor} catalogue with the SkyBoT predicted positions within a $3\sigma$ search radius, where $\sigma$ is given by:
\begin{equation}
{\sigma}^2={\sigma_{WTS}}^2+{\sigma_{SBT}}^2
\end{equation}

Then, we cross-matched the list of asteroid counterpart candidates with our {\it Non-moving Source Catalogue}. This way, we removed potential mismatches with other celestial sources or image artifacts. This was especially important for those cases in which the large uncertainties in the estimated SkyBoT's asteroid position led to very wide search radii. On the contrary, after visually checking the results, we realized that for some asteroids the uncertainties in the coordinates provided by SkyBoT ($\sigma_{SBT}$) based on osculating elements, were underestimated, which led to  too small search radii. Thus, to avoid losing asteroid detections, we set a minimum search radius of 3\arcsec. 

\begin{figure}
        \centering
        \includegraphics[width=\hsize]{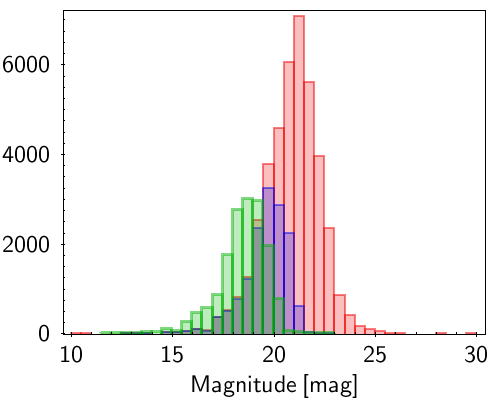}
        \caption{Distribution of apparent magnitudes from SkyBoT of the 41\,804 expected asteroids in the WTS (visual magnitudes, in red), the final sample of detected asteroids with the Sighted method (visual magnitudes, in blue) and the final sample of detected asteroids with the Sighted and Blind (see Section~\ref{blind.method}) methods (J-band magnitudes, in green).}
         \label{fig.histV}
\end{figure}

According to SkyBoT, there were 41\,804 potential asteroid detections in the studied set of J-band images. For 25\,589 of them, we did not find any asteroid counterpart.
Several reasons could have prevented us from identifying the asteroid counterpart, mainly the asteroid brightness to be below the image detection limit, as shown in Figure~\ref{fig.histV}
It displays the distributions in apparent visual magnitude of the 41\,804 expected asteroids and of the final sample of asteroids detected with the Sighted method, and the distribution in the J-band of the final sample of asteroids detected with the Sighted and Blind (see Section~\ref{blind.method}) methods.
Typical magnitude limit in the survey is around 18.5\,mag in the $J$ band, although it varies among images and nights. Also slow motion asteroids would be misidentified as non-moving sources, and defects in the images or the asteroid counterpart blending with a field star could explain the missing asteroid detections.

For another 15\,508 of them, we obtained a unique asteroid counterpart candidate for each potential asteroid detection by SkyBoT. For the remaining 707 potential asteroid detections, we found 3\,433 asteroid counterpart candidates. The large number of counterparts per SkyBoT's asteroid position can be ascribed mainly to the large orbital uncertainties provided by SkyBoT (hence, large search radii), but also to noisy images or image artifacts due to the presence of a nearby bright saturated star.


\begin{figure}
        \centering
        \includegraphics[width=\hsize]{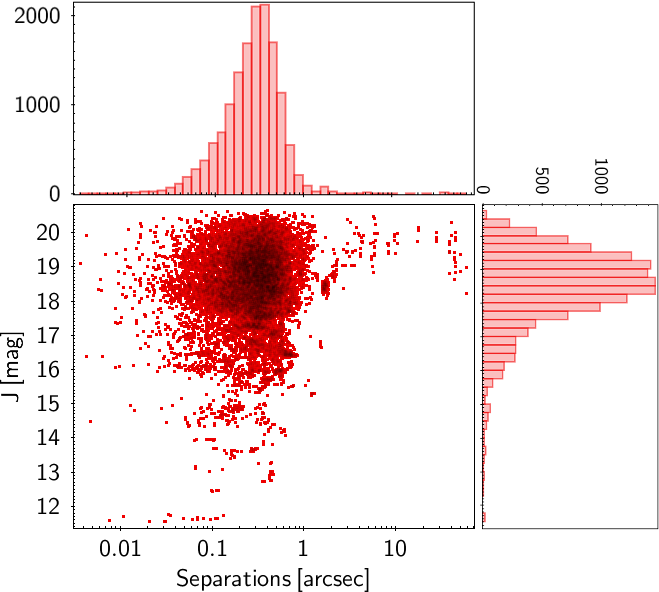}
        \caption{Apparent $J$ band magnitudes vs. separations between the SkyBoT and the counterpart candidates positions of the Sighted method. Distributions of separations (on top) and magnitudes (on left) are shown.}
         \label{fig.sepmag}
\end{figure}

Of the 15\,508 potential asteroid detections with unique asteroid counterpart candidates, 315 correspond to 315 asteroids with only one detection in the complete survey. They were automatically discarded since no validation test could be applied for them.
The other 15\,193 potential asteroid detections correspond to 1\,713 different asteroids with counterpart candidates at different epochs. These multi-epoch asteroid detection allowed us to define an additional selection criteria. For each asteroid, the separation between the position of the asteroid in the WTS image and the position predicted by SkyBoT tends to be similar from one image to another. We therefore rejected any asteroid detection counterpart deviating by more than $3\,\sigma_{sep}$ from the typical separation between the asteroid counterpart candidates and the predicted SkyBoT positions for this asteroid, with $\sigma_{sep}$ being the standard deviation of these separations. Additionally, since 94\% of the sample showed $\sigma_{sep}$\,$\leq$\,0.3\arcsec, we visually inspected the counterpart candidates of any asteroid with $\sigma_{sep}$\,$>$\,0.3\arcsec, and removed any suspicious counterpart. In total, 676 asteroid counterpart candidates were rejected.


We also checked whether any of the 707 potential asteroid detections with multiple asteroid counterpart candidates  corresponded to any of the 1\,713 asteroids previously studied. For each asteroid in common, we run the same test than before using the typical separation and its corresponding $\sigma_{sep}$ obtained in the previous exercise. This way, we were able to recover asteroid counterpart candidates for another 188 SkyBoT's possible detections. The rest of the potential asteroid detections with multiple counterparts per SkyBoT position was neglected. Thus, we got a unique asteroid counterpart candidate for 14\,705 SkyBoT asteroid position.



Separations between celestial coordinates provided by SkyBoT and those extracted in the WTS images for the asteroid counterpart candidates are shown in Figure~\ref{fig.sepmag} together with their apparent $J$ magnitudes.
The distribution of separations peaks at 0.3\arcsec, although a few of them could reach separations larger than 10\arcsec. For 90\% of the sample, the separation is under 0.58\arcsec. 
These numbers reflect the high accuracy in the orbital parameters available at SkyBoT for most of the sample. They also illustrate the capability of our procedure to identify asteroids, as well as the high efficiency of our method to find asteroids with poorly determined orbits (i.e., those with a high uncertainty on their orbital osculating elements, turning into a high uncertainty on sky positions computed by SkyBoT, see \citealt{Des13}).

We performed an additional test based on colours on the 14\,705 asteroid counterpart candidates. Asteroids are visible because they reflect the light of the Sun. Hence, their typical $V-J$ colour lies in the 1.2 -- 1.7\,mag range \citep{Holm06}. In addition, the amplitude of a typical light-curve of an asteroid could be as large as 0.6\,mag \citep{Harris16}. Thus, we visually inspected all asteroid detections with $V-J$ out of the colour range between 0.6 and 2.3\,mag, where V is the predicted magnitude estimated by SkyBoT and J is the magnitude that we obtained from the WTS images. This other check allowed us to identify 152 mismatches and 15 asteroid counterpart candidates affected by saturation problems. All of them were removed from the sample. Thus, the sample was reduced to 14\,538 asteroid counterpart candidates.


Finally, we were able to measure proper motions and their errors from 14\,181 detections of 1\,591 asteroids. To do that, we considered only positions and epochs within the same night. Hence, single asteroid detections over a night will not have proper motion measurements. 

There is a correlation between the temporal coverage of the observations and the proper motion errors obtained from them: as expected, the smaller the time baseline, the larger the proper motion error is. In our sample, the time baseline spans from 2 minutes to more than 8 hours. The 89\% of calculated proper motions have errors smaller than 1\arcsec/h in both components (right ascension and declination), which correspond to time baselines longer than $\sim$0.5\,h. For shorter temporal coverage, the errors increase with the time baseline up to 17\arcsec/h in each component for the shortest temporal intervals.

\begin{figure*}
        \centering
        \includegraphics[width=0.45\textwidth]{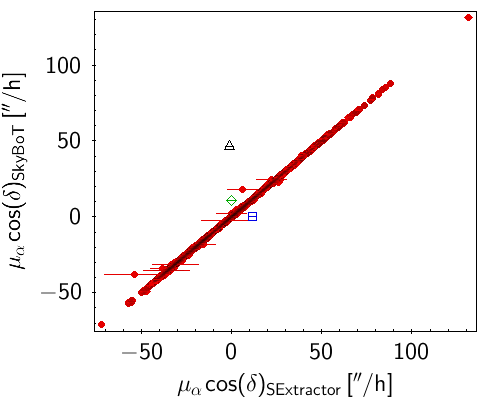}
        \includegraphics[width=0.45\textwidth]{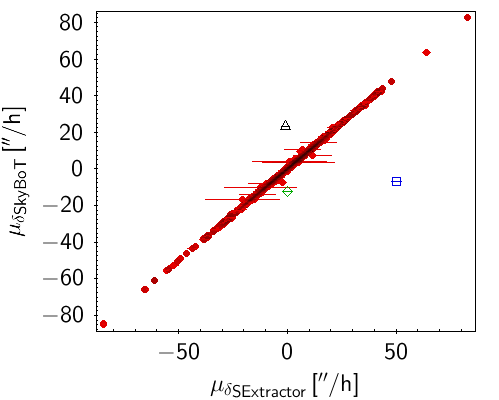}
        \caption{Comparison of proper motions provided by SkyBoT and those computed in the Sighted method. Errors from SkyBoT and {\sc SExtractor} are computed from the standard deviation of proper motions over the same night. Vertical errorbars are smaller than the used symbol. Horizontal errorbars correspond to the uncertainties in {\sc SExtractor} proper motions. Open symbols in both panels stand for the asteroids 2005 EE135 (blue square), 2008 RY130 (green diamond) and 2015 BY519 (black triangle).}
         \label{fig.pmcomp}
\end{figure*}

Our measured proper motions and those provided by SkyBoT are compared in Figure~\ref{fig.pmcomp}. An excellent agreement was reached with 
only three exceptions: 
2005 EE135, 2008 RY130, and 2015 BY519. These asteroids show differences in the total proper motions between 16 and 50\arcsec/h. The proper motions of the three asteroids were obtained from only two measurements separated by $\sim$16 minutes in time. In the first case (2005 EE135), the error in the predicted position provided by SkyBoT is the largest in the whole catalogue ($>$\,65\arcsec), and the counterpart candidates were found at more than 3\arcmin\ from the predicted position. In the second case (2008 RY130), the error in the predicted sky position is also large (16\arcsec), and the counterpart candidates were found at near that separation. 
The third asteroid (2015 BY519) has smaller error in the predicted position provided by SkyBoT ($\sim$\,6.5\arcsec), but it was detected at near 3$\sigma_{SBT}$ (17\arcsec). After a visual inspection of the six images involved, we confirmed that these detections were not related to the asteroid suggested by SkyBoT but to faint field sources. Hence, we discarded these asteroid counterpart candidates and the final number of asteroid detections found with this method reached 14\,532 of 1\,615 asteroids.
Their magnitude distribution in the $J$ band is shown in Figure~\ref{fig.sepmag}. It peaks at 18.45\,mag, near the detection limit of the survey.

Table~\ref{tab.summary_sighted} summarizes the numbers of discarded or recovered asteroid detections in each step performed during this method, for clarity.

\begin{table}
        \centering
        \caption {Summary of the number of asteroid detections discarded (--) or recovered (+) in the Sighted method. }
        \label{tab.summary_sighted}
        \begin{tabular}{l c l} 
        \hline \hline
        \noalign{\smallskip}
Status  &   Number of potential &  Comment \\	
        &   asteroid detections &    \\  

        \noalign{\smallskip}
        \hline
        \noalign{\smallskip}
        \noalign{\smallskip}
  &   41\,804 &   From SkyBoT \\
--  &   25\,589     & Without counterpart in {\sc SExtractor}   \\
--  &   707        &   With too many counterparts in {\sc SExtractor}    \\
+   &   188        &   Recovered with $\sigma_{sep}<0.3$\arcsec    \\
--  &   315    &   With only one detection in the survey    \\
--  &   676    &   Discarded with $\sigma_{sep}>0.3$\arcsec  \\
--  &   152    &   $V-J$ out of range  \\
--  &   15     &   Saturated detections    \\
--  &   6      &   Mismatches with faint field sources    \\

\noalign{\smallskip}    
        \hline
 \noalign{\smallskip}   
 &  14\,532 &   Valid asteroid detections  \\
 \noalign{\smallskip}   
 \hline
        \end{tabular}
\end{table}


\subsection{Blind method: ssos Pipeline}\label{blind.method}

To cross-check the positions recovered with the Sighted method described above and to additionally search for unknown asteroids, we further applied the \textit{ssos} pipeline to the WTS images \citep{Mahlke2019n}. \textit{ssos} is a versatile tool to detect and identify Solar System objects in astronomical images. It does not require prior knowledge, e.g. from SkyBoT queries, and therefore allows for the detection of both known and unknown asteroids. A drawback is that each object has to be observed at least three times during a single night in order to be recovered, as the apparent motion cannot be reliably evaluated based on only one or two detections. We give a brief outline of the detection principles here. For details, the reader is referred to the online documentation\footnote{\url{https://ssos.readthedocs.io}}.

\subsubsection{Image cleaning}

We used the same 30\,558 calibrated J-band images as outlined above in Sect.\,\ref{sec:img_clean}, including the 770 images with problems with the third detector. The applied cuts in pixel space differed slightly from the previous ones. We aimed to include as much of the image edges as possible, as each single asteroid detection increases the probability of the asteroid being identified by the \textit{ssos} pipeline. Thus, we applied cuts of 23 pixels on all four image edges and the same cut as before at the upper left corner ($y > 2.1624x$+$1934.34$). Furthermore, we added a cut to exclude an ubiquitous imaging artifact in the bottom right corner, at $y < 420$ for $x > 2000$.

The detection method of the \textit{ssos} pipeline is based on the identification of moving sources by comparing their positions from image to image. This is done on a nightly basis, to ensure that the associated positions belong to the same asteroid. To describe the apparent motion of a source, at least three detections in one night are needed. Thus, in a first step, the input images were grouped by observation night and by the celestial region they cover.
The 30\,558 images were divided into 790 observation groups, with varying amounts of images composing each group. Of them, six groups were discarded as they consisted of fewer than 3 observations. In total, 30\,548 images in 784 groups were searched.

\subsubsection{Source catalogue and photometric calibration}

The 30\,548 images were reduced to source catalogues using {\sc SExtractor}. Image artifacts and low signal-to-noise sources could prevent the detection of an asteroid from its apparent motion by the \textit{ssos} pipeline. In order to avoid this as much as possible, we decreased the amount of source deblending and used a slightly more homogeneous background estimation than with the Sighted method. The {\sc SExtractor} parameters used are given in Table~\ref{table.sexparam} in Appendix~\ref{app.config}. 

The same photometric calibration as derived in Sect.\,\ref{phot} was used for the asteroid positions found with \textit{ssos}.

\subsubsection{Asteroid identification}

The following analysis was then run independently on the source catalogues obtained by {\sc SExtractor} and grouped by observing nights and sky regions, as described above.

The sources detected in each image were associated to each other by overlaying the detections in the celestial-coordinates-space. This way, different detections of the same non-moving source will be close to each other (separations should be similar to the seeing) and then detections will be associated to a single source. The same applied to image artifacts such as bad pixels, and to moving sources where consecutive detections were too close in the sky. 
The astrometric solution of the source catalogues of the same night and celestial region was computed by SCAMP \citep{2006ASPC..351..112B}. The SCAMP configuration parameters can be found in Table~\ref{table.scampparam} in Appendix~\ref{app.config}.

Now, the source catalogues were compared to a reference catalogue. The SDSS-DR12 catalogue \citep{Alam15} was chosen as the reference catalogue for three of the WTS regions, and the {\it Gaia}-DR2 catalogue for the fourth region at 29\arcdeg\ of ecliptic latitude. Although the SDSS-DR12 was deeper than the {\it Gaia}-DR2, the former did not cover this last region. 

SCAMP associated the detections of transient sources within a cross-match radius between consecutive detections. The cross-match radius was dictated by the mean image cadence, and it was set to 15\arcsec\ after a trail-and-error iteration of several test groups and visual inspection of the output samples. If set too small, fast asteroids will not be detected as the single source positions move out of the cross-match radius.
If set too large, SCAMP will associate random detections of other sources (artificial or astrophysical) to the asteroid, resulting in non-linear apparent motion and a subsequent rejection due to the pipeline algorithm (see \citealt{mahlke2018mining} for details). 

At this point of the pipeline, the source catalogues consisted on the possible asteroid candidates, imaging artifacts, and astrophysical sources not present in the reference catalogues. As next and final step of the \textit{ssos} pipeline, a set of configurable filter algorithms aims to separate asteroids from the remaining sources. The main characteristic that we used to differentiate asteroids was their linear apparent motion. In a first step, all sources with only one or two detections were removed, as their motion cannot be judged reliably. Afterwards, we applied a minimum proper motion cut of 0.8\arcsec/h, equivalent to a source moving from one pixel to the next one within one hour. While there may be distant Kuiper-belt objects with apparent motion rates below this cut, they are likely too faint to be recovered and the trade-off to remove contaminants present in the source catalogues is more valuable. 

We then performed linear least-squares fits on the sources' right ascension and declination over observation epoch. Sources displaying linear motion in both dimensions and with a goodness-of-fit parameter $R^2$\,$>$0.9 were accepted. 

The final output of the \textit{ssos} pipeline is a catalogue of asteroid candidate detections. We first cross-matched it with the SkyBoT database within 20\arcsec\ and without imposing any restriction to the position errors of the asteroids, and removed outliers in proper motion space analogous to the Sighted method (see Sect.\,\ref{sec:ast_ind}). Furthermore, we extracted cutout images of all positions and visually inspected them to remove possible contaminants, which were mostly diffraction spikes around bright stars and other persistent image artifacts. While visual inspection was time-consuming, it gave us a large degree of confidence in the asteroid candidates, which is especially valuable when claiming the discovery of unknown objects.

The final sample contained 1\,165 distinct asteroids, with a total of 9\,897 detections. Of them, 654 correspond to 182 potentially unknown asteroids, i.e. without counterparts within a 20\arcsec\ in the SkyBoT database.

\section{Results and discussion}
\label{sec:results}

\subsection{Method comparison}\label{sec:method_comp}

Two different, but complementary, approaches were used to search for asteroids in the WTS images. Depending on the configuration of the different steps carried out with each method, the output samples had varying degrees of completeness and purity. Here, we compare the results from both methods and join both catalogues in a final one.

\begin{table}
        \centering
        \caption {Comparison of the number of asteroids and asteroid positions recovered with the Sighted ($S$) and Blind ($B$) methods. }
        \label{tab.summary}
        \begin{tabular}{l c c} 
        \hline \hline
        \noalign{\smallskip}
Method & Number of positions	&	Number of asteroids   \\	

        \noalign{\smallskip}
        \hline
        \noalign{\smallskip}
        \noalign{\smallskip}
$S$   &   14\,532 &   1\,615     \\
\noalign{\smallskip}   
$B$   &   9\,897  &   1\,165    \\
\noalign{\smallskip}   
$S$ \& $B$    &   8\,768  &   954   \\
\noalign{\smallskip}   
$S$ NOT $B$ &   5\,764  &   661  \\
        \noalign{\smallskip}   
$B$ NOT $S$ &   1\,129  &   401    \\
\noalign{\smallskip}    
        \hline
 \noalign{\smallskip}   
 $S$ + $B$  &   15\,661 &   1\,821    \\
 \noalign{\smallskip}   
 \hline
        \end{tabular}
\end{table}

The Sighted method, using prior information from the SkyBoT database, recovered 14\,532 positions of 1\,615 distinct asteroids. The Blind approach, based on the \textit{ssos} pipeline, detected 1\,165 asteroids at 9\,897 positions.

The comparison of the results obtained with both methods reveals that the Sighted method is able to identify 5\,764 asteroid positions that the Blind method does not recover. This is due to several reasons, for example, a misclasification during the detection-association step by SCAMP. Also the asteroids being too faint to be detected during the source extraction process (note that the {\sc SExtractor} settings for the Blind method aimed towards reducing spurious detections, which naturally removed fainter asteroids from the output sample), or the asteroids being not visible in more than two images during a single night, render them invisible to the Blind method.
On the other hand, the linear fit basis of the \textit{ssos} pipeline allows the identification of 1\,129 asteroid positions not recovered by the Sighted method. Of them, 654 are associated to 182 new asteroids, impossible to be recovered by the Sighted method since there is no related information in SkyBoT and, thus, the service could not provide predicted positions for them.

Table~\ref{tab.summary} summarizes, for clarity, the numbers here presented. $S$ and $B$ stand for Sighted and Blind method, respectively.
The number of asteroids indicated in the table must be considered as a reference. Note that we are comparing the number of asteroid positions instead, since the same asteroid could have been detected by the Sighted and Blind methods in equal or different positions. It could therefore appear counted twice in the table (except in the numbers accounting for $S + B$).

If we consider the results obtained by the Blind method as a reference because of its more restrictive conditions, we observe that 89\% of the asteroid positions detected by this method were also identified in the Sighted method. The differences in sky positions of common detections are below 0.4\arcsec\ for the vast majority of them. Only in 10 cases, the differences went up to 1\arcsec. This occurred towards the images edges, where the distortion is larger. SCAMP, used by \textit{ssos}, can correct this distortion after deriving the astrometric solution, which was not corrected by the Sighted method. It would explain these large differences in sky positions.

The large percent of common detections found by both methods and the number of additional positions obtained by each of them, strongly validate each other and highlight their complementary performance.

It is not possible to carry out an unbiased comparison of these results with those obtained using large-area surveys like KiDS \citep{mahlke2018mining} or VISTA VHS \citep{Popescu16,Popescu18}, or with those specifically dedicated to SSO monitoring (e.g., \citealt{vaduvescu2018}). The main reason lies upon the differences in the used instrumentation, which lead to different magnitude and detection limits, and the observing strategy: while surveys like KiDS or VISTA VHS cover large areas in the sky, transit surveys like WTS focus in small areas but provide time-resolved information.

\subsection{The asteroid sample}\label{sec:ast_sample}

\begin{figure}
   \centering
   \includegraphics[width=\hsize]{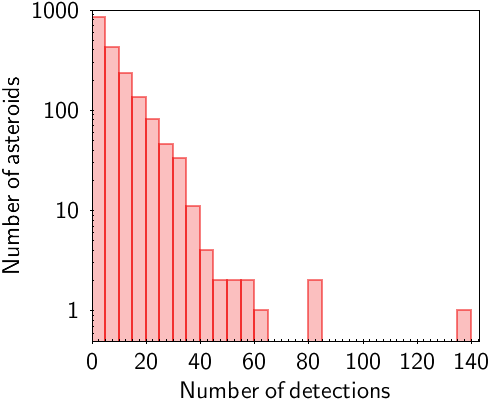}
      \caption{Distribution of the number of detections per asteroid for all known and new discovered asteroids in the sample.}
         \label{fig.detec_p_ast}
   \end{figure}

Finally, we ended up with a total of 15\,661 detections corresponding to 1\,821 different asteroids. The distribution of the number of detections per asteroid is shown in Figure~\ref{fig.detec_p_ast}. Half of the asteroids presents five or less detections, while 90\% have under 20 detections in the survey.

For all of them, we provide $\alpha$ and $\delta$ coordinates, $\mu_{\alpha} \cos(\delta)$ and $\mu_{\delta}$ proper motions (see Sect.\,\ref{pm}), the observing epoch, and the obtained $J$ band magnitude. This information can be gathered from {\em The SVO archive of asteroids} at the Spanish Virtual Observatory portal \footnote{\url{http://svo2.cab.inta-csic.es/vocats/v2/wtsasteroids/}} (see Appendix~\ref{app.svocat}). 

The number of asteroids and detections found in the different WTS fields is presented in Table~\ref{tab.regions}.

\subsection{Dynamical classes}\label{sec:dynamical_class}

Table~\ref{tab.class} summarizes the number of asteroids and detections in the sample sorted by asteroid dynamical class, as provided by SkyBoT.
Asteroid population is classified in SkyBoT as defined by the international community according to the characteristics of their orbits\footnote{\url{http://vo.imcce.fr/webservices/skybot/?documentation}}.

\begin{table}
        \centering
        \caption {Number of asteroids and detections per asteroid dynamical class.} 
        \label{tab.class}
        \begin{tabular}{l r r} 
        \hline \hline
        \noalign{\smallskip}
Dynamical  & \# asteroids &		\# asteroid     \\
class		&  (fraction)			&	 detections	\\

        \noalign{\smallskip}
        \hline
        \noalign{\smallskip}
        \noalign{\smallskip}
Main Belt: Middle	&	640	(35.1)	&	5\,968\\  
Main Belt: Outer	&	530	(29.1)	&	4\,865	\\ 
Main Belt: Inner	&	322	(17.7)	&	2\,936	\\ 
Hungaria	&	52 ~~(2.9)	&	376	\\  
Trojan	&	42 ~~(2.3)	&	499\\  
Mars-Crosser	&	26 ~~(1.4)	&	201\\ 
Main Belt: Cybele	& 11 ~~(0.6)	&	78	\\  
Main Belt: Hilda	&	8 ~~(0.4)	&	60\\ 
Near-Earth Asteroid: Apollo	&	4 ~~(0.2)	&	10\\ 
Near-Earth Asteroid: Amor	&	4 ~~(0.2)	&	14\\ 
Unknown  & 182 (10.0) & 654 \\

\noalign{\smallskip}    
        \hline
        \end{tabular}
\end{table}

\begin{figure}
   \centering
   \includegraphics[width=\hsize]{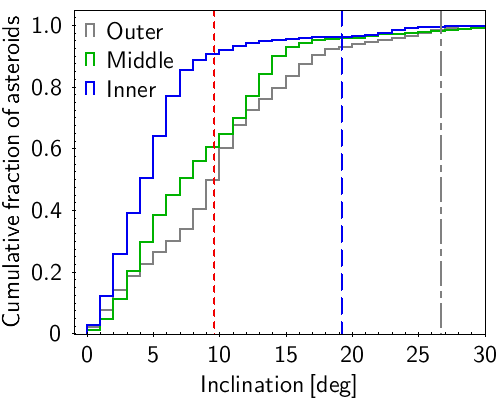}
      \caption{Cumulative distribution of the orbital inclinations of Inner (blue), Middle (green) and Outer (gray) Main Belt asteroids taken from the ASTORB database. Vertical dashed lines indicate the ecliptic latitudes of three WTS observed fields: -9.59\arcdeg\ (red), +19.19\arcdeg\ (blue), and +26.69\arcdeg\ (gray).}
         \label{fig:inclinations}
\end{figure}

The number of detected Inner Main Belt asteroids is near 50\% and 40\% lower than the number of Outer and Middle Main Belt asteroids detected, respectively. Due to the closer distance to Earth and higher albedos (i.e., brighter apparent magnitudes) of Inner Main Belt asteroids, we would expect to have detected a larger fraction of them.
We compare these numbers with the numbers of Inner, Middle and Outer Main Belt asteroids found when cross-matching SkyBoT to our images in a 0.17\arcdeg\ radius. The number of Inner Main Belt asteroids lying in our images is half the number of Middle and Outer Main Belt asteroids, separately.
This difference could be explained by the different orbital inclinations observed in Inner, Middle and Outer Main Belt asteroids. Figure~\ref{fig:inclinations} shows the cumulative distribution of the inclinations of these three classes of Main Belt asteroids taken as a reference from the ASTORB database. We notice that in the observed field by the WTS at ecliptic latitude -9.59\arcdeg, 90\% of the Inner Main Belt asteroids present lower inclinations than the field's latitude. The percentage of Middle and Outer Main Belt asteroids with lower inclinations is $\sim$50\% at the same latitude. This would translate into a larger probability of finding Middle and Outer Main Belt asteroids compared to Inner ones.
Also the larger apparent motion of Inner Belt asteroids would prevent us from detecting them consecutively within the targeted field and night. 


\begin{figure}
   \centering
   \includegraphics[width=\hsize]{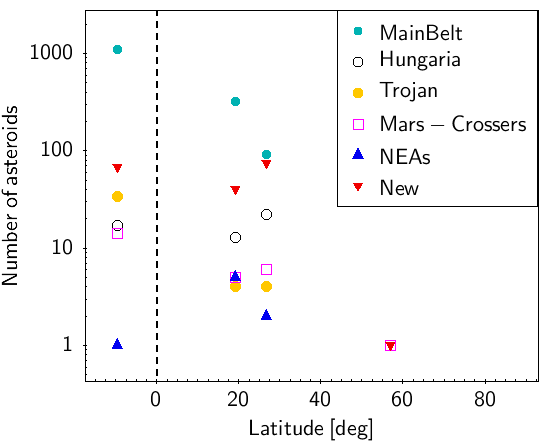}
      \caption{Number of asteroids of each dynamical class as a function of the latitude, which absolute values can be used for comparison with the orbital inclination. Green filled circles stand for Main Belt asteroids, black open circles for Hungaria objects, yellow filled circles for Jupiter's Trojans, magenta open squares for Mars-Crossers, blue filled up triangles for NEAs and red filled down triangles for new discovered asteroids.}
         \label{fig.numdist_perast}
   \end{figure}

Figure~\ref{fig.numdist_perast} shows the number of asteroids of each dynamical class detected in each WTS field, including new discoveries. The absolute value of the latitude would act as a proxy of the orbital inclination. The field that lies farthest from the ecliptic only counts with a Mars-Crosser asteroid and one new discovered asteroid detected five and four times, respectively.

\subsection{Asteroid families}\label{sec:astfam}

Asteroid families, created by collisions, are a unique tool to study both the internal structure of their parent body, and the surface and dynamical evolution of their member, hence of the asteroid belt \citep{demeo19, spoto15, osz15}.
The Asteroids--Dynamic Site\footnote{\url{https://newton.spacedys.com/astdys2/index.php?pc=0}} (AstDyS-2) provides catalogues of the proper elements of the asteroids as well as of asteroid families and family membership obtained from the analysis of proper elements \citep[see][]{astfam1}.

In our catalogue, 410 asteroids out of the 1\,639 known asteroids are associated to 44 different families. 
Of them, 168 are core family members identified by Hierarchical Clustering Method, 217 are members added by attribution to core families and 25 are members of small or satellite families. They are distributed along the Hungaria, Main Belt and Trojan populations.
While the majority of the families identified here contains less than 20 asteroids from our sample each, there are three families with more than 60 members detected in this work: the ones of Vesta, Eunomia and Eos (family codes 4, 15 and 221, respectively).
We include the related family code in our catalogue.

\begin{figure}
   \centering
   \includegraphics[width=\hsize]{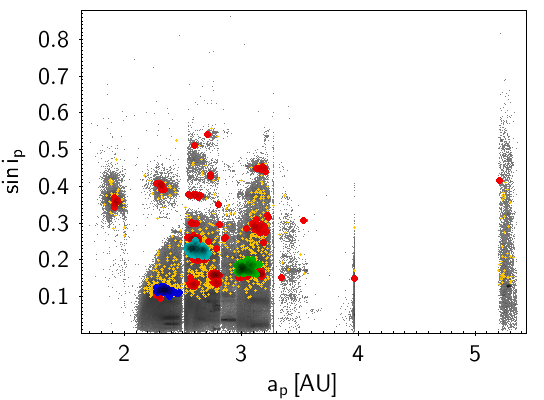}
      \caption{Sine of proper inclination versus proper semi-major axis of all asteroids in the AstDyS-2 database (light gray dots), asteroids in our sample not linked to any asteroid family (yellow pluses) and asteroids in our sample associated to an asteroid family (filled red circles). Filled dark blue, light blue and green circles represent the asteroids in the Vesta, Eunomia and Eos families, respectively.}
         \label{fig.fam}
   \end{figure}

The distribution in the proper elements space $\sin{i_p}$ vs. $a_p$ is shown in Figure~\ref{fig.fam}. Asteroids in families are represented with a different symbol and colour. Those of Vesta's, Eunomia's and Eos' families are highlighted.

\subsection{Magnitudes}\label{sec:magnitudes}

The distribution of calibrated J-band magnitudes of the 15\,661 asteroid counterparts identified in this work is shown in green in Figure~\ref{fig.histV}. It peaks at around 18.9\,mag, which reflects the completion limit of our search. Note that this limit is consistent with and slightly higher than the approximated magnitude limit obtained for the survey at 18.5\,mag.

\begin{figure}
   \centering
   \includegraphics[width=\hsize]{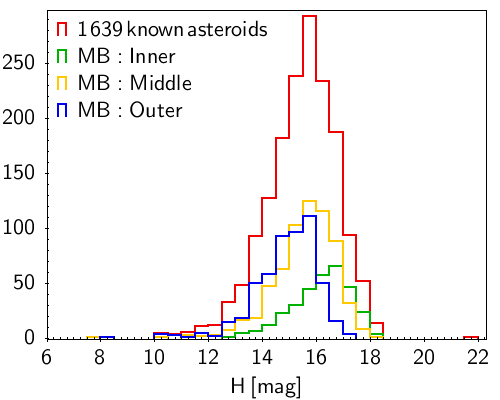}
   \caption{Distribution of absolute magnitude H of the 1\,639 known asteroids in the sample, and of Inner, Middle and Outer Main Belt asteroids.}
         \label{fig.mags}
   \end{figure}

Figure~\ref{fig.mags} shows the distribution of absolute magnitudes H of the 1\,639 known asteroids in the sample. H magnitudes were taken from the Asteroids--Dynamic Site. The mean absolute magnitude H of detected known asteroids in the sample is 15.4\,mag.

The distribution of absolute H magnitudes of Inner, Middle and Outer Main Belt asteroids detected in the survey is also shown in Figure~\ref{fig.mags}. Owing to the closer distance to Earth of Inner Main Belt asteroids compared with Middle and Outer's, and different albedo on average \citep{masiero11, demeo13}, we were able to detect fainter Inner than Middle or Outer Main Belt objects. There is therefore a global observational bias that is also reflected in our distribution.

\subsection{Proper motions}\label{pm}

We measured total proper motions for 1\,794 asteroids. They span from 0.5\arcsec/h to 142\arcsec/h. This is, from slow Main Belt asteroids and Jupiter's Trojans to fast near-Earth asteroids.

As an average, asteroids with measured proper motions move at 24.3\arcsec/h. Main Belt asteroids dominate the distribution with an average apparent motion of 23.7\arcsec/h. As expected, Jupiter's Trojans are the slowest objects in the sample with a mean total proper motion of 17.3\arcsec/h, while NEAs show the fastest mean apparent motion of 73.8\arcsec/h.

\subsection{Shape models}

The backup nature of the WTS provides observations randomly distributed over time. Therefore, the number of detections per asteroid (Figure~\ref{fig.detec_p_ast}) is typically too small and the observing geometry too narrow to contain enough information about the shape and spin state of individual asteroids. Therefore, we combined our data with photometry from  the Lowell Observatory photometry database, which contains re-calibrated sparse-in-time photometry from large sky surveys observed between 2000--2012 \citep{Osz.ea:11,Bow.ea:14}. The combined data set may lead to unique shape/spin solution even when the individual data sets (WFCAM and Lowell) are not sufficient alone. Similarly to previous works \citep{Han.ea:13b, Han.ea:16, Dur.ea:16, Dur.ea:16c, Dur.Han:18}, we looked for the best-fit model by using the light-curve inversion method of \cite{Kaa.ea:01} and scanning the parameter space on an interval of periods 2--100\,h with ten initial poles for every period. Then we selected the formally best solution with the lowest $\chi^2$ fit. We checked if the minimum in $\chi^2$ was significantly deeper than all other solutions and if the corresponding shape model had the rotation axis aligned with the maximum principal axis of inertia tensor. 

\begin{figure}
        \centering
        \includegraphics[width=\hsize]{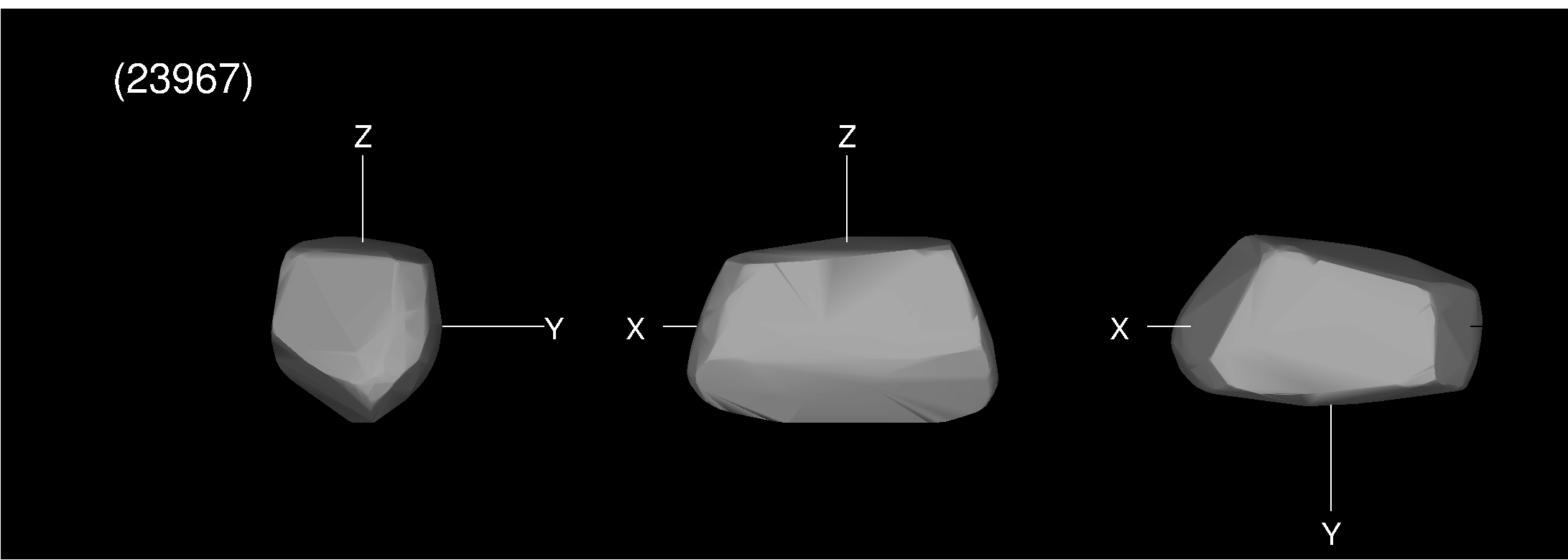}
        \includegraphics[width=\hsize]{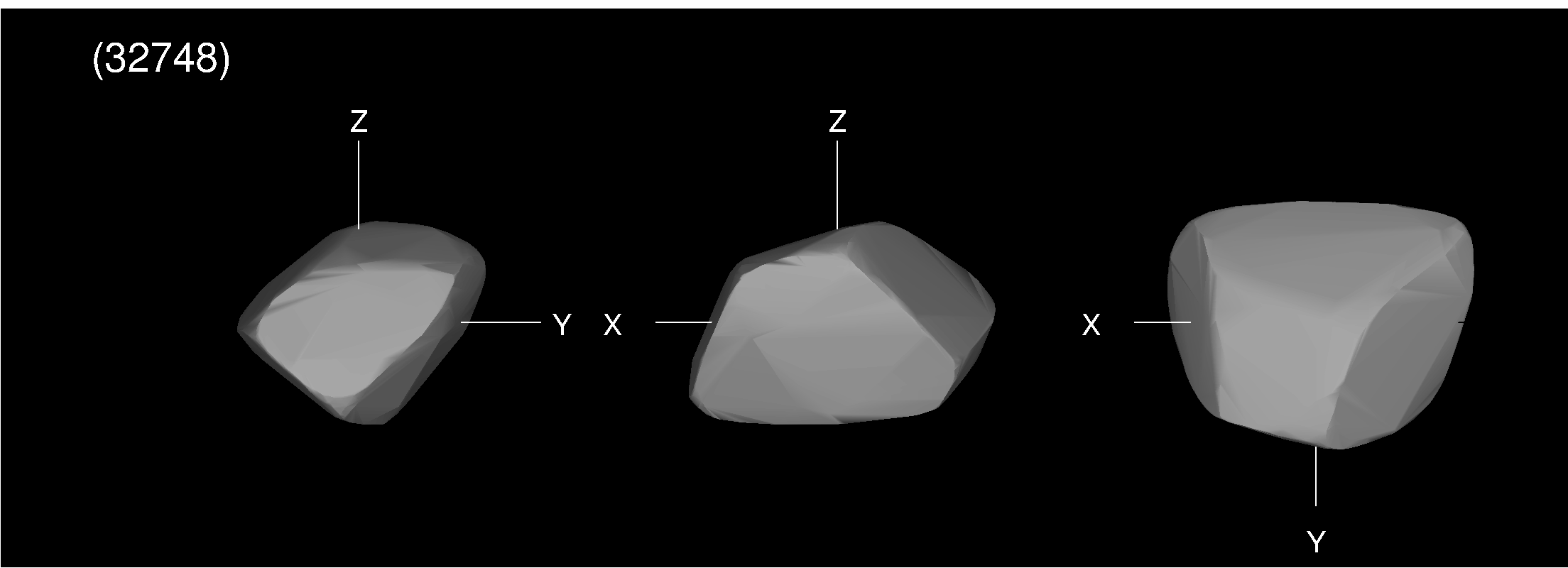}
        \includegraphics[width=\hsize]{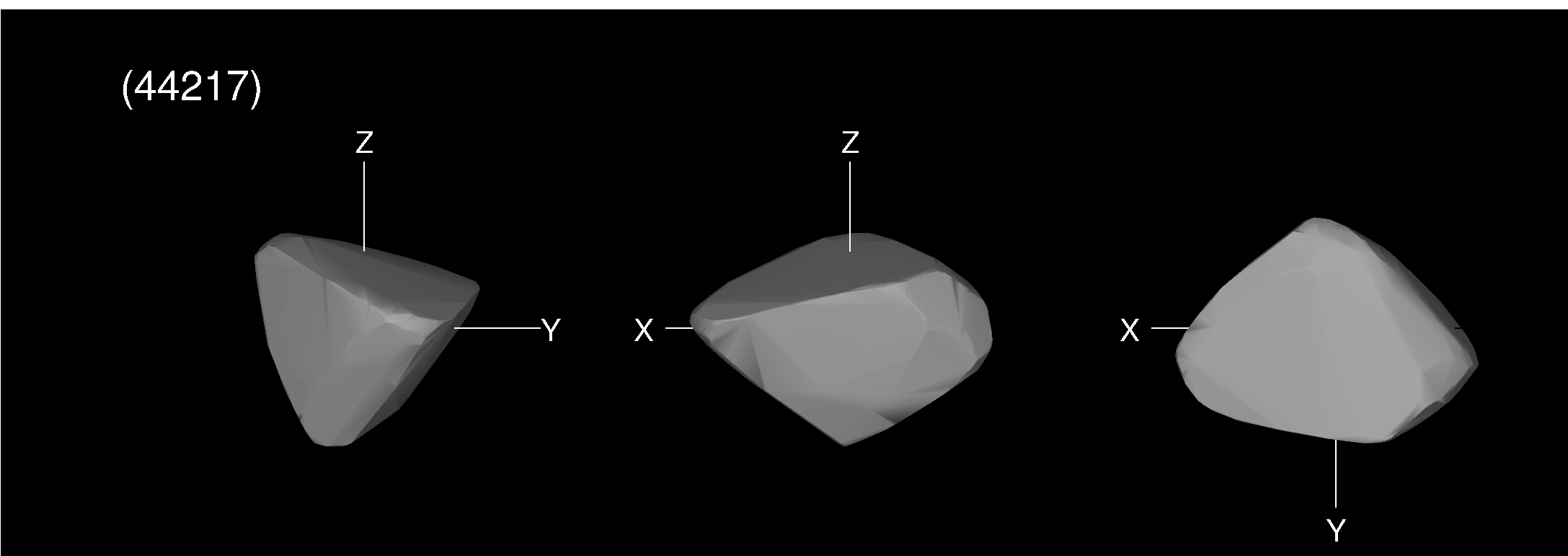}
        \includegraphics[width=\hsize]{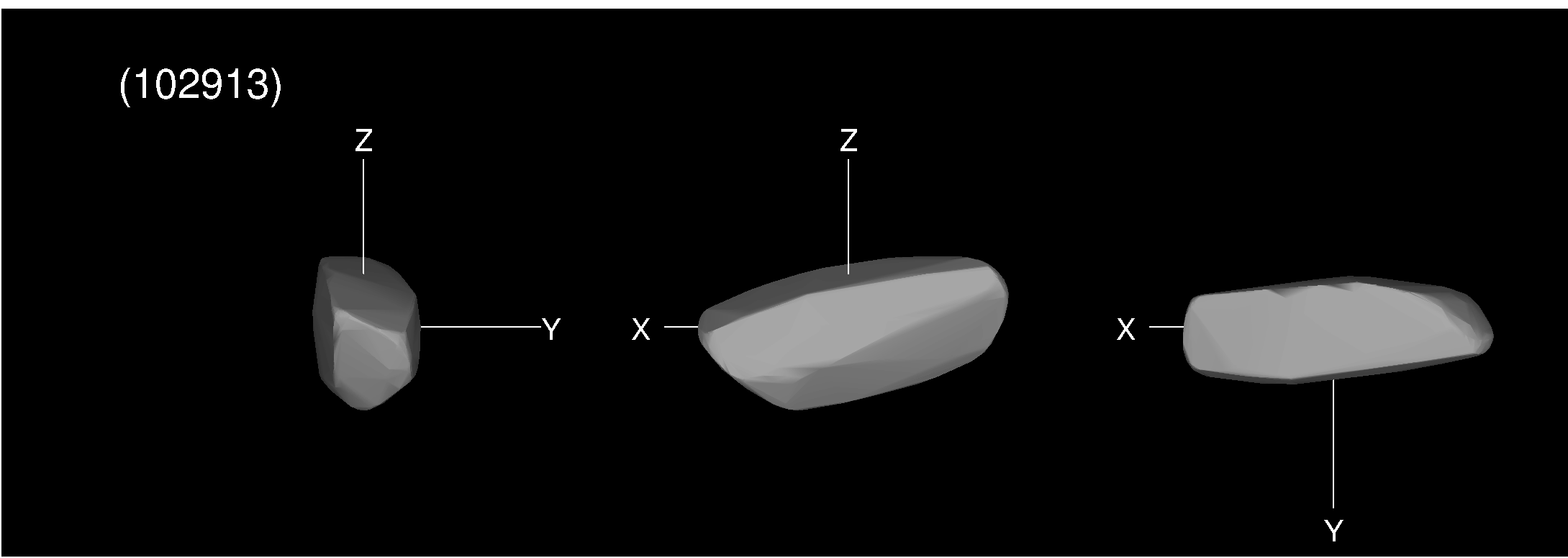}
        \includegraphics[width=\hsize]{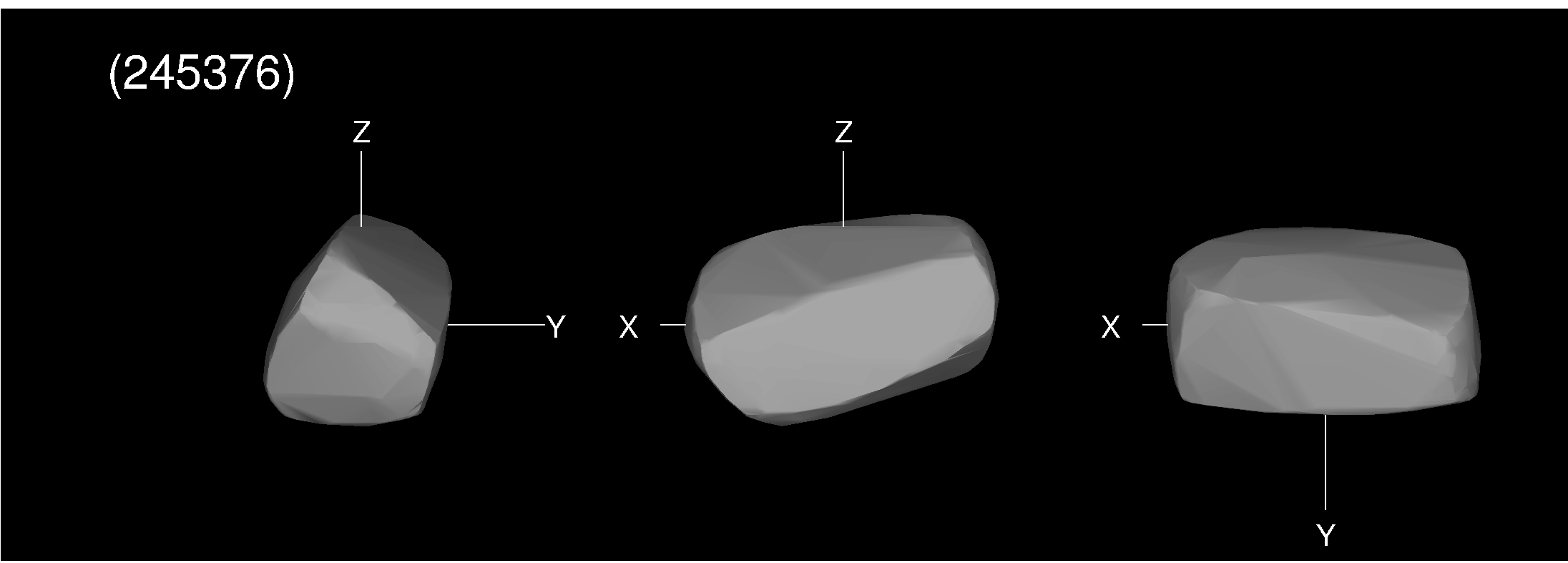}
        \caption{Convex shape models of asteroids from Table~\ref{tab:models} corresponding to the first pole solution. $Z$ is the rotation axis.}
        \label{fig:models}
  \end{figure}

For most of the asteroids in our sample, even the combination of Lowell and WFCAM photometry was not sufficient to provide a unique shape and spin model.
Usually, there were many solutions with different rotation periods or spin axis directions that provided essentially the same fit to the data. The only five asteroids that passed all the reliability tests are listed in Table~\ref{tab:models}. These are the new shape/spin models that we derived from WFCAM and Lowell data, with periods ranging from 5.5 to 16.0\,h. The uncertainty in the rotation period is of the order of the last decimal place given in Table~\ref{tab:models}(between $1\mathrm{e}{-4}$ and $1\mathrm{e}{-5}$ hours), the uncertainty in pole direction is around 10--20\arcdeg. 
Figure~\ref{fig:models} shows the convex shape models of the five asteroids in Table~\ref{tab:models}.
For asteroid (44217) Whittle, we have an independent confirmation that our spin solution is correct. \cite{Gal:09} observed this asteroid in 2009 and derived the light-curve period of $9.7 \pm 0.1$\,h, which is consistent with our value of $9.78271$\,h.
For (23967) 1998 XQ12, we have now a full shape and spin state model with two possible spin solutions. Spin parameters of our new model agree with the rotation period and pole ecliptic latitude of a partial model derived by \cite{Durech18} from Lowell photomery (the same set as we used) and WISE thermal light-curves.
The folded light-curves of both of them are shown shown in Figure~\ref{fig:light_curve} in Appendix~\ref{app.fig}. The spread of the WTS data of the remaining three asteroids in Table~\ref{tab:models}, prevents us from obtaining a good shaped light-curve and are therefore not shown in the figure.

In principle, to maximize the scientific output of WFCAM asteroid photometry, one could combine the data not only with Lowell Observatory photometry, but with all photometry available in archives -- light-curves archived in the minor planet light-curve database \citep[LCDB][]{War.ea:09}, photometry from Palomar Transient Survey \citep{Was.ea:15}, Gaia \citep{Spo.ea:18}, etc. However, this is out of the scope of this paper.

   \begin{table*}[t]
   \centering
   \caption{List of new asteroid models. For each asteroid, we list one or two pole directions in the ecliptic coordinates $(\lambda, \beta)$, the sidereal rotation period $P$, the rotation period from LCDB $P_\mathrm{LCDB}$ (if available) and its quality code $U$, which is an assessment of the quality of period solution \citep{War.ea:09}. $U = 2$ means that the result was based on less than full coverage of the light-curve.}
   \label{tab:models} 
    \begin{tabular}{c l r r r r @{} d @{} d l}
      \hline\hline
      \multicolumn{2}{c}{Asteroid}		& \multicolumn{1}{c}{$\lambda_1$}	& \multicolumn{1}{c}{$\beta_1$}	& \multicolumn{1}{c}{$\lambda_2$}	& \multicolumn{1}{c}{$\beta_2$}	& \multicolumn{1}{c}{$P$}	& \multicolumn{1}{c}{$P_\mathrm{LCDB}$}	& \multicolumn{1}{c}{U} \\
      Number	& name/designation		& \multicolumn{1}{c}{[deg]}		& \multicolumn{1}{c}{[deg]}	& \multicolumn{1}{c}{[deg]}		& \multicolumn{1}{c}{[deg]}	& \multicolumn{1}{c}{[h]}	& \multicolumn{1}{c}{[h]} 		& 			\\
      \hline
 23967 & 1998 XQ12  &   190 & $-18$ &   357 & $-45$ &  5.47479 &    5.47479\pm0.00002      &    2       \\
 32748 & 1981 EY7   &   107 &    29 &   276 &    49 &  9.53678 &             &       \\
 44217 & Whittle    &    63 &    51 &   216 &    34 &  9.78271 & 9.7\pm0.1   &     2 \\
102913 & 1999 XT21  &   201 & $-23$ &       &       &  15.9591 &             &       \\
245376 & 2005 GC79  &     6 &    33 &       &       &  7.93897 &             &       \\
      \hline
    \end{tabular}  
   \end{table*}
    
\subsection{Identification of NEO candidates among unknown asteroids}
10\,\% of the recovered asteroids could not be matched to a known object using SkyBoT. Determining their orbital class and possibly uncovering NEAs among them could be a valuable effort. However, these 182 objects only comprise 654 detections. The majority of them was only observed three times within several hours, rendering a sound characterization impossible. Nevertheless, in the following, we attempt to identify NEOs in the sample using both the NEO Rating Tool\footnote{\url{https://minorplanetcenter.net/iau/NEO/PossNEO.html}} of the Minor Planet Center and the EURONEAR NEA Checker\footnote{\url{http://www.euronear.org/tools/NEACheck.php}}.

\begin{figure}
    \centering
    \includegraphics[width=0.5\textwidth]{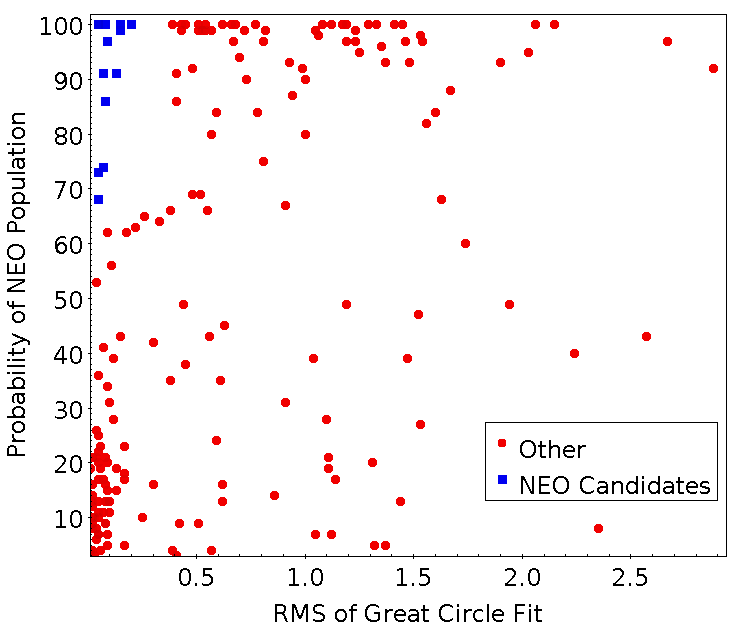}
    \includegraphics[width=\hsize]{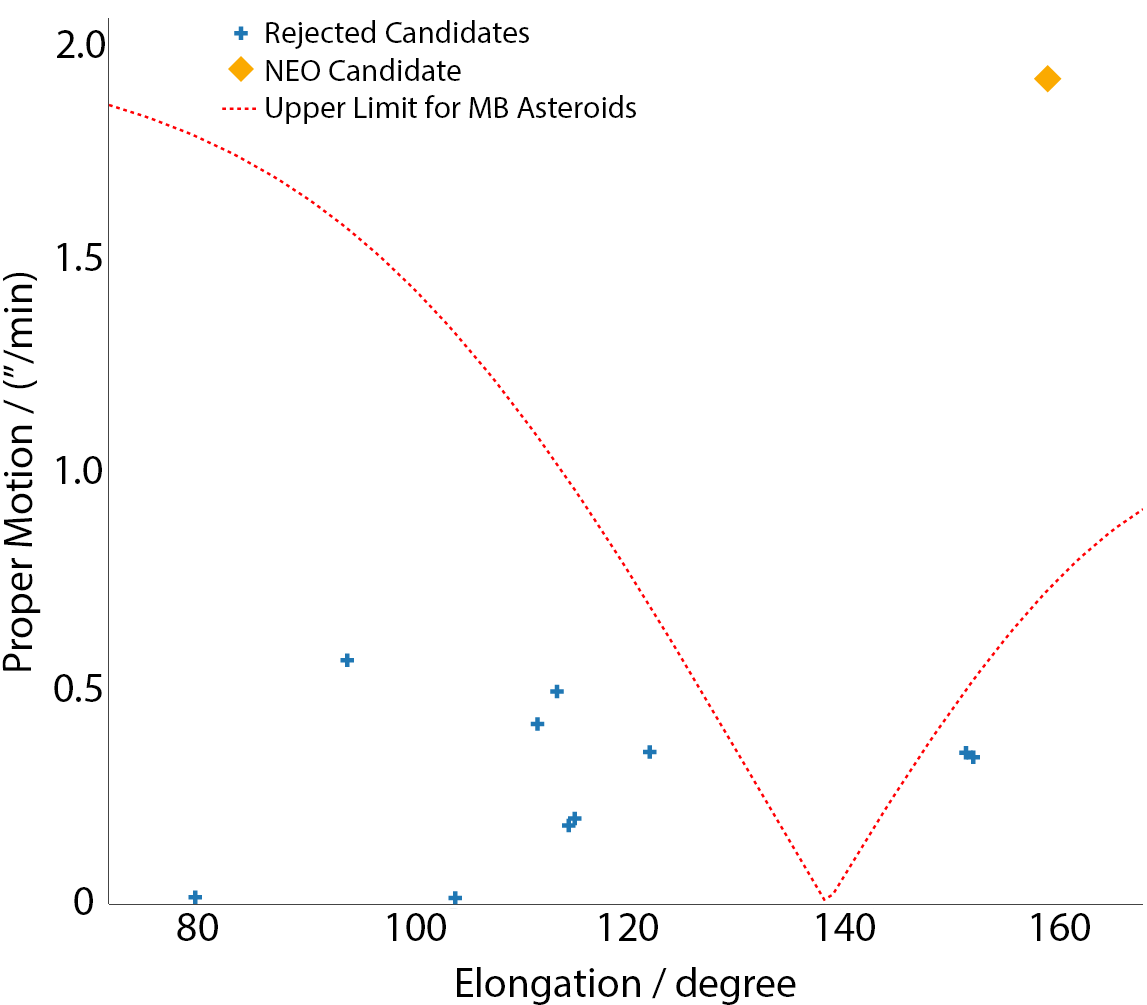}
    \caption{{\it Top panel}: The probability of the unknown asteroids belonging to the NEO population against the RMS of a great-circle-fit to their observations, as calculated by the MPC' NEO Rating Tool. Blue squares mark objects with RMS errors below 0.2\,\arcsec and NEO scores above 65. Red circles depict the remainder of the unknown asteroid sample.
    {\it Bottom panel}: The upper limit of the proper motion of Main Belt asteroids in relation to their solar elongation as derived by the EURONEAR NEA Checker is shown by the red, dotted line. The blue dots mark the proper motion and solar elongation of the 12 NEO candidates identified with the MPC's NEO Rating Tool. One candidate is clearly above the derived proper motion limit.}
    \label{fig:neo_vs_rms}
\end{figure}

In the top panel of Figure~\ref{fig:neo_vs_rms}, we depict the probability of the unknown asteroids belonging to the NEO populations against the RMS errors. It is apparent that the fraction of asteroids with large NEO scores is unreasonably high for this random sample of asteroids. Furthermore, 60\,\% of the objects depicts RMS errors larger than 0.2\,\arcsec. As these objects have been visually confirmed to be asteroids and the sample of known asteroids has not displayed observational flaws, we account the large RMS and general overestimation of NEO scores to the short arc of the observations. Few detections per asteroid covering a period of hours are likely not sufficient for a meaningful orbit estimation.

Nevertheless, when focusing on the objects with RMS below 0.2\,\arcsec, we can observe a cluster of objects with small NEO scores, which are likely Main Belt objects. In the same region, 12 objects show probabilities above 65\%, a threshold defined by the MPC as worthy for follow-up observations. In a next step, we applied the NEA Checker by the EURONEAR project to these 12 NEO candidates. Briefly, the NEA Checker assumes circular asteroid orbits co-planar to the orbit of Earth and derives upper limit for the proper motion of Main Belt asteroids against the solar elongation of the objects at the epoch of observation \citep{2011P&SS...59.1632V}. Bottom panel in Figure~\ref{fig:neo_vs_rms} shows the proper motion limit for Inner Main Belt according to this model, and the proper motions and solar elongations of the 12 NEO candidates.

We see that the NEO score of 11 candidates was overestimated and that they are consistent with Main Belt objects. One candidate shows a proper motion that exceeds the limit by a factor of almost three. While additional observations will be necessary to confirm, this object is likely in the near-Earth domain. The three observations of this candidate were acquired over seven minutes and are depicted in Figure~\ref{fig:neo_candidate} in Appendix~\ref{app.fig}. It displayed a proper motion of 97\,\arcsec/h.

\subsection{Submission to Minor Planet Center}

The asteroid positions identified in this work have been reported to the MPC.
Of the 15\,661 asteroid positions reported, 15\,246 corresponding to 1\,789 distinct asteroids have been accepted, including the 182 potential new discoveries. It represents 97.4\% of the sample.

Of the reported asteroid positions, 415 were rejected by the MPC. They correspond to 162 different asteroids. Note that not all the reported positions of these 162 asteroids were rejected. We did not obtain the cause of rejection from the MPC but we try to deduce it here. For 271 out of the 415 asteroid positions, there was already an entry in the database at the same or very close epoch (within a day). We assume that this is the reason for rejection. Another 112 positions show entries in the MPC database at epochs as close as 10.8\,d and as far as several months from the epoch of our positions. Therefore, a close-in-time detection does not explain rejection. 
On the other hand, the MPC documentation states that a batch containing any single positions will be returned entirely. We noticed that seven asteroids with 32 positions in total contain at least a single night detection, and therefore, have been rejected.
Hence, we conclude that, of the submitted positions, 0.9\% (112+32 asteroid positions) did not pass the MPC filters for acceptance and near 1.7\% (271 asteroid positions) was rejected for reasons not related to the quality of our data.

\section{Conclusions}
\label{sec:conclusions}

In this work, we presented two different methods to search for asteroids serendipitously observed in archive images. One of the methods (Sighted method) relies on the orbital osculating elements for already known asteroids available from the SkyBoT service. The other method (Blind method) looks for moving objects in consecutive images without any prior knowledge. Both methods were validated in the set of J-band images from the WFCAM Transit Survey collected in the framework of the RoPACS project. The two methods have demonstrated to be very efficient in the detection of asteroids, with a large percentage ($\sim$89\%, 8\,768 asteroid positions) of common detections.
Besides, each of them contributes with additional 5\,764 (Sighted method) and 1\,129 (Blind method) asteroid positions.
Both methods are then complementary: while the Sighted method is more effective in detecting faint asteroids and asteroids with less than three detections per night, the Blind method is able to discover new asteroids.

Joining together the results of both methods we built a catalogue which is publicly available at the Spanish Virtual Observatory portal (see Appendix~\ref{app.svocat}). The catalogue contains 15\,661 detections of 1\,821 asteroids, including 182 potential new discoveries. The magnitude distribution of our detections peaks around 18.9\,mag in the J-band, which reflects the completion limit of our search. This information was submitted to the Minor Planet Center to improve the orbital parameters of the known asteroids and report the new ones.

Of the sample of detected known asteroids in the survey, 410 are classified into 44 asteroid families.
In addition, over 80\% of the detected known asteroids belong to the Main Belt and only 0.4\% are NEAs.
Of the 182 potential new discoveries, only one is likely in the near-Earth domain, belonging the rest of them to the Main Belt population. The fraction of NEAs in the sample of discovered asteroids is comensurable to the fraction obtained among known asteroids.

The non-uniform distribution of observations over time due to the nature of the WTS programme, prevented us from obtaining complete light-curves of the observed SSOs.
We therefore combined our photometric data with those at the Lowell Observatory to derive shape/spin models for the detected known asteroids. We found unique solution for five of our asteroids. For the rest of asteroids, the number of detections and the quality of the combined data did not allow us to get a unique solution.

The effectiveness and complementarity of the two methodologies followed in this work for the detection of asteroids in the images of the WFCAM Transit Survey has been proved. They can, in fact,  be applied to similar searches in other large-area astronomical surveys.

\section*{Acknowledgements}
This research has been financed by ASTERICS, a project supported by the European Commission Framework Programme Horizon 2020 Research and Innovation action under grant agreement n. 653477. This research has made use of the Spanish Virtual Observatory (http://svo.cab.inta-csic.es) supported from the Spanish MINECO/FEDER through grant AYA2017-84089-P.
M.C.C. and F.J.E. acknowledge financial support from the Tec2Space-CM project (P2018/NMT-4291).
S.B.F. acknowledge support by the Spanish State Research Agency (AEI) through project No. ESP2017-87676-C5-1-R and No. MDM-2017-0737 Unidad de Excelencia ''Mar\'ia de Maeztu''-Centro de Astrobiolog\'ia (CSIC-INTA).
This research has made use of Topcat \citep{Taylor05} and STILTS \citep{Taylor06}. This research has made use of "Aladin sky atlas" developed at CDS, Strasbourg Observatory, France. This research has made use of IMCCE's SkyBoT VO tool \citep{Ber06}. The work of J\v{D} was supported by the grant 18-04514J of the Czech Science Foundation. MM is funded by the European Space Agency under the research contract C4000122918.
MCC and FJE acknowledge financial support from the Tec2Space-CM project (P2018/NMT-4291).

+
\bibliographystyle{mnras}
\bibliography{mnras.bib}



\clearpage

\appendix
\onecolumn

\section{Configuration parameters}\label{app.config}

\clearpage
\begin{table}
        \centering
        \caption {Configuration parameters for {\sc SExtractor}.}
        \label{table.sexparam}
        \begin{tabular}{l r r}
        \hline \hline
        \noalign{\smallskip}
{\sc SExtractor}	&	Sighted method 	& Blind method	\\ 
parameter	&	value 	&	value	\\ 
        \noalign{\smallskip}
        \hline
        \noalign{\smallskip}
        \noalign{\smallskip}
        
DETECT\_TYPE  &    CCD &    CCD    	\\
DETECT\_MINAREA &  4    &  6    	\\
THRESH\_TYPE  & RELATIVE & RELATIVE \\
DETECT\_THRESH  &  2    &  2      	\\
FILTER         &  Y   & Y          	\\
FILTER\_NAME    &  default.conv  &  default.conv 	\\
DEBLEND\_NTHRESH&  32   & 16     	\\
DEBLEND\_MINCONT&  0.005 & 0.01    	\\
CLEAN        &    Y     & Y        	\\
CLEAN\_PARAM   &   1.0     & 1.0  	\\
MASK\_TYPE    &    CORRECT  & CORRECT    	\\
 \noalign{\smallskip}
  \noalign{\smallskip}
WEIGHT\_TYPE  &   --  &  BACKGROUND \\
PHOT\_APERTURES  & 5.0  &    -- 	\\   
PHOT\_AUTOPARAMS & 2.5, 3.5 & 1.5, 0.8 	\\  
PHOT\_PETROPARAMS& 2.0, 3.5 &  --	\\  
PHOT\_AUTOAPERS & --  & 0.0,0.0 \\
SATUR\_LEVEL   &   50000.0  &   50000.0	\\  
SATUR\_KEY     &   SATURATE &  SATURATE 	\\  
MAG\_ZEROPOINT  &  0.0  &   0.0    	\\  
MAG\_GAMMA     &   4.0  &   4.0    	\\  
GAIN          &   0.0   &  0.0    	\\  
GAIN\_KEY      &   GAIN &  GAIN     	\\  
PIXEL\_SCALE    &  0.251 &  0    	\\  
 \noalign{\smallskip}
  \noalign{\smallskip}
SEEING\_FWHM    &  0.75 & 1. 	\\
STARNNW\_NAME   &  default.nnw & default.nnw	\\
\noalign{\smallskip}
 \noalign{\smallskip}
BACK\_TYPE    &  AUTO & AUTO \\
BACK\_SIZE      &  64 & 128 	\\
BACK\_FILTERSIZE & 3  & 3	\\
BACKPHOTO\_TYPE &  GLOBAL & GLOBAL 	\\
 \noalign{\smallskip}
  \noalign{\smallskip}
MEMORY\_OBJSTACK & 3000 &3000	\\
MEMORY\_PIXSTACK & 300000 &300000	\\
MEMORY\_BUFSIZE &  1024 &1024	\\
\noalign{\smallskip}
        \hline
        \end{tabular}
\end{table}


\begin{table}
        \centering
        \caption {Configuration parameters for {\sc SCAMP}. Only the parameters which could affect the outcome are listed.}
        \label{table.scampparam}
        \begin{tabular}{ll}
        \hline \hline
        \noalign{\smallskip}

ASTREF\_CATALOG      &   SDSS-R12           \\
DEFAULT      &   DEFAULT           \\
ASTREFMAG\_LIMITS      &   -99.0,99.0    \\
MATCH      &   Y               \\
MATCH\_NMAX      &   0               \\
PIXSCALE\_MAXERR      &   1.2             \\
POSANGLE\_MAXERR      &   5.0             \\
POSITION\_MAXERR      &   1.0           \\
MATCH\_RESOL      &   0               \\
MATCH\_FLIPPED      &   N               \\
MOSAIC\_TYPE      &   UNCHANGED       \\
FIXFOCALPLANE\_NMIN      &   1             \\
\noalign{\smallskip}
\hline
\noalign{\smallskip}
CROSSID\_RADIUS      &   15.0           \\
\noalign{\smallskip}
\hline
\noalign{\smallskip}
SOLVE\_ASTROM      &   Y               \\
PROJECTION\_TYPE      &   SAME            \\
STABILITY\_TYPE      &   INSTRUMENT      \\
DISTORT\_GROUPS      &   1,1             \\
DISTORT\_DEGREES      &   2               \\
FOCDISTORT\_DEGREE      &   1               \\
ASTREF\_WEIGHT      &   1.0             \\
\hline
\noalign{\smallskip}
ASTRACCURACY\_TYPE      &   SIGMA-PIXEL \\
ASTRACCURACY\_KEY      &   ASTRACCU        \\
ASTR\_ACCURACY      &   0.01            \\
ASTRCLIP\_NSIGMA      &   3.0             \\
COMPUTE\_PARALLAXES      &   N               \\
COMPUTE\_PROPERMOTIONS   &   Y               \\
CORRECT\_COLOURSHIFTS    &   N               \\
INCLUDE\_ASTREFCATALOG   &   Y               \\
ASTR\_FLAGSMASK          &   236          \\
ASTR\_IMAFLAGSMASK       &   0x0             \\
\hline
\noalign{\smallskip}
SOLVE\_PHOTOM            &   N               \\
SN\_THRESHOLDS           &   1.5,100.0      \\
FWHM\_THRESHOLDS         &   0.0,1000.0       \\
ELLIPTICITY\_MAX         &   1.0             \\
FLAGS\_MASK              &   236          \\
WEIGHTFLAGS\_MASK        &   0x00ff          \\
IMAFLAGS\_MASK           &   0x0            \\      
\noalign{\smallskip}
        \hline
        \noalign{\smallskip}
        \noalign{\smallskip}

\end{tabular}
\end{table}

\clearpage
\section{Additional figures}\label{app.fig}

\clearpage
\begin{figure}
        \centering
        \includegraphics[width=0.45\hsize]{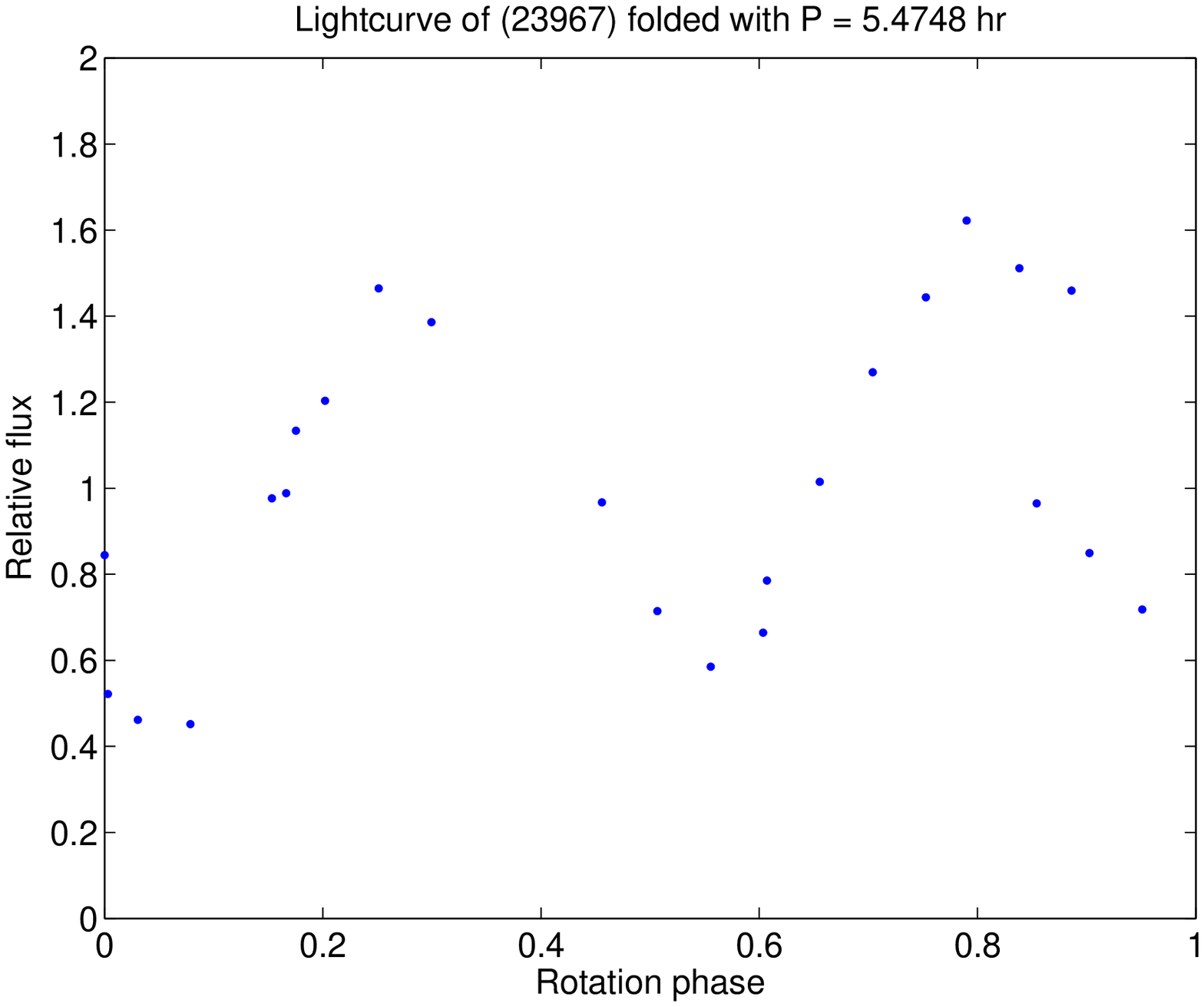}
        \includegraphics[width=0.45\hsize]{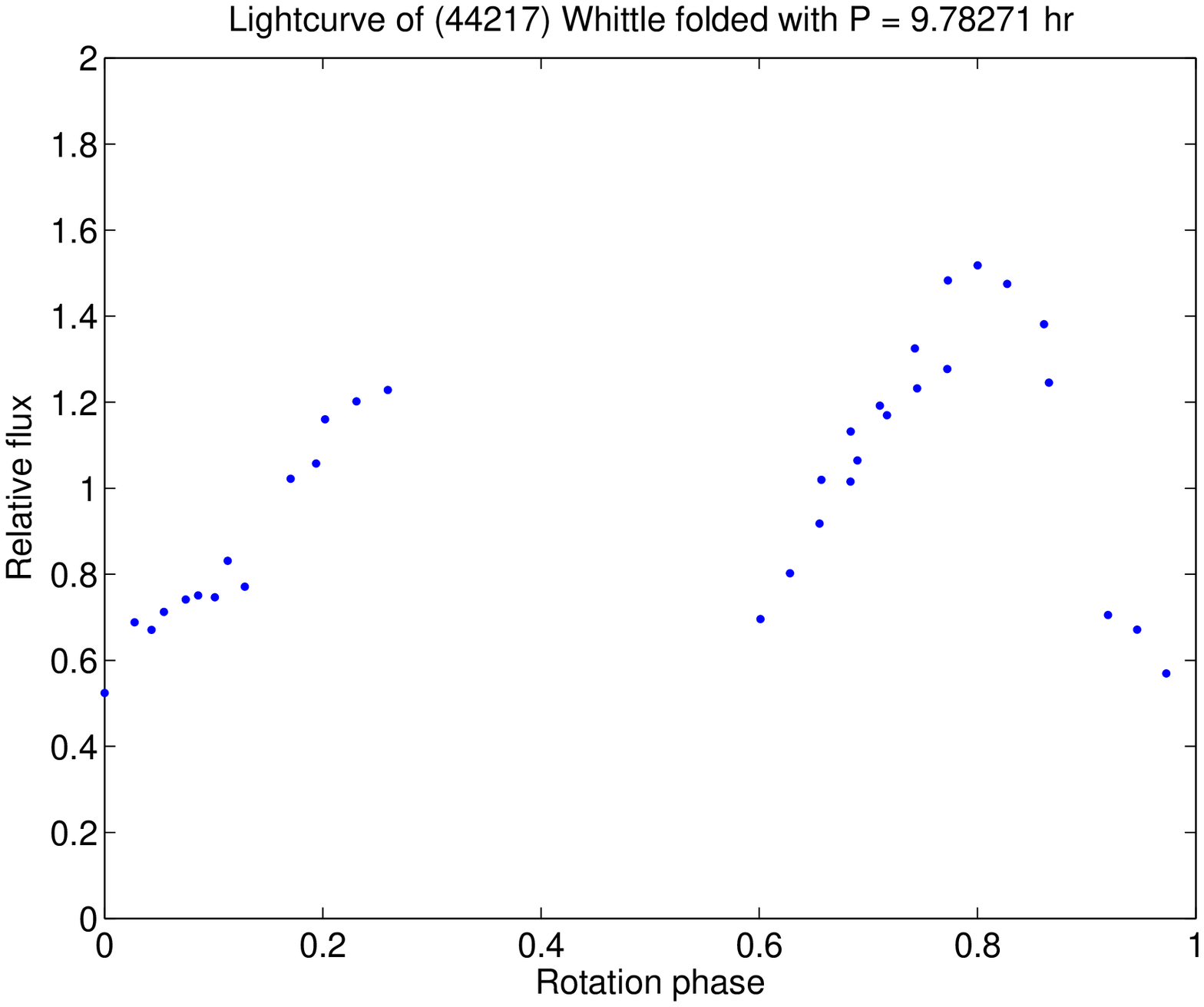}
        \caption{Composite light-curve of the asteroids 23967 and 44217 folded with respect to the obtained period in Table~\ref{tab:models}.}
         \label{fig:light_curve}
  \end{figure}

 \begin{figure*}
     \centering
     \includegraphics[width=1\textwidth]{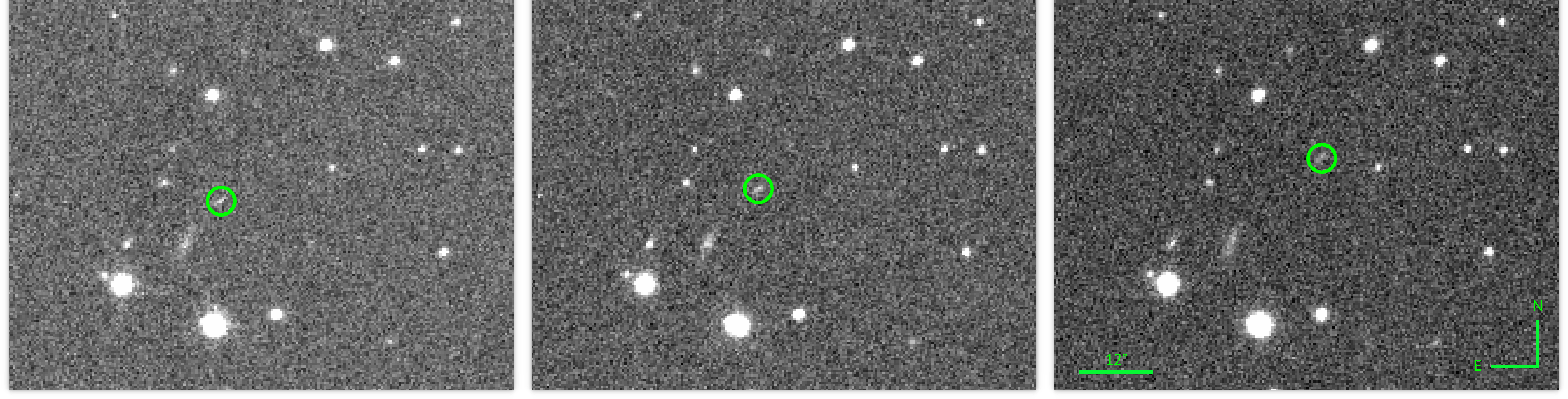}
     \caption{The three observations of the NEO candidate identified from the unknown sample of asteorids using the MPC NEO Rating Tool and the EURONEAR NEA Checker. The asteroid is surrounded by a green circle in the three images and had a proper motion of 97\,\arcsec/h at the time of observation. The time difference between the first (left) and second (middle) observation is 130\,s, the one between the second and third (right) 322\,s. In the last frame, the travelled distance of the asteroid is indicated in the bottom left, equaling about 12\,\arcsec, while on the bottom right, the north-east alignment of the frame is indicated, with north being towards the top.}
     \label{fig:neo_candidate}
 \end{figure*}

\clearpage
\section{Online catalogue service}\label{app.svocat}

In order to help the astronomical community on using the catalogue of asteroids identified in the WTS images, we have developed an archive system that can be accessed from a webpage\footnote{\url{http://svo2.cab.inta-csic.es/vocats/v2/wtsasteroids/}} or through a Virtual Observatory ConeSearch\footnote{e.g. \url{http://svo2.cab.inta-csic.es/vocats/v2/wtsasteroids/cs.php?RA=53.612&DEC=39.150&SR=0.1&VERB=2}}.

\begin{figure*}
        \centering
        \includegraphics[width=\hsize]{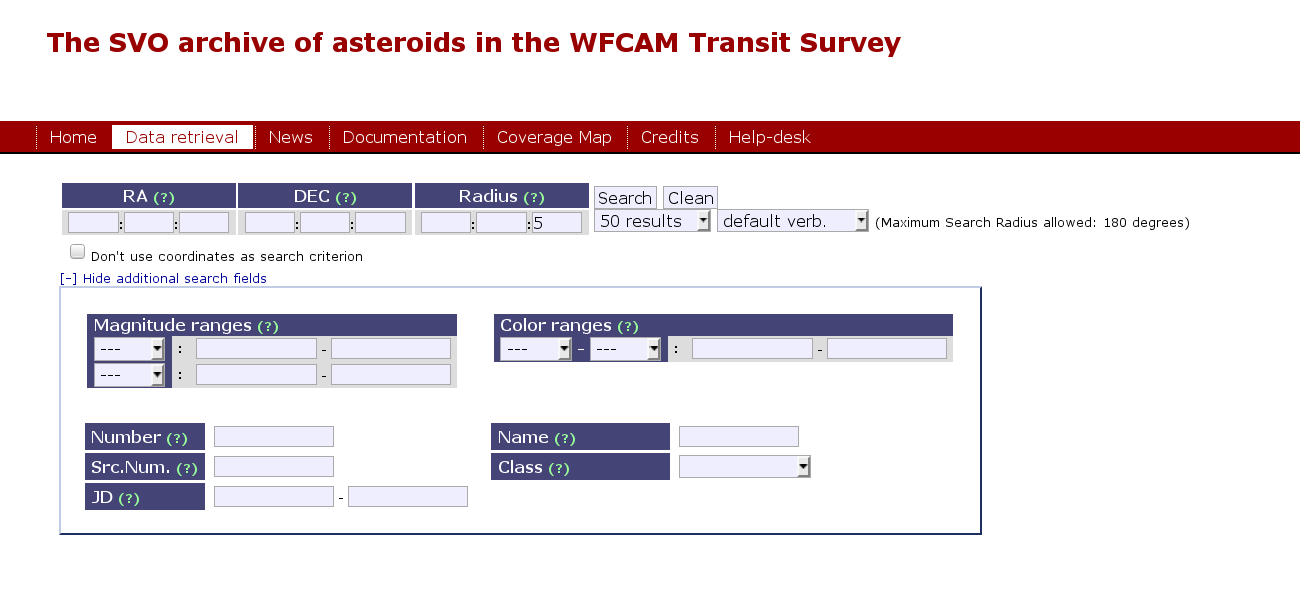}
        \caption{Screenshot of the archive search interface that permits simple queries.}
         \label{fig.svoquery}
\end{figure*}

The archive system implements a very simple search interface (see Figure~\ref{fig.svoquery}) that permits queries by position, number, name, asteroid dynamical class, and range of magnitudes, colors or epochs. The user can also select the maximum number of sources to return (with values from 10 to unlimited). The system also implements a link to the latest version of Asteroid Orbital Elements Database in VizieR, which is a copy of the Asteroid Observing Services from the Asteroid Database at Lowell Observatory\footnote{\url{https://asteroid.lowell.edu/main/}}.

The result of the query is a HTML table with all the sources found in the archive fulfilling the search criteria. The result can also be downloaded as a VOTable or a CSV file. Detailed information on the output fields can be obtained placing the mouse over the question mark (``?") located close to the name of the column. The archive also implements the SAMP\footnote{\url{http://www.ivoa.net/documents/SAMP/}} (Simple Application Messaging) Virtual Observatory protocol. SAMP allows Virtual Observatory applications to communicate with each other in a seamless and transparent manner for the user. This way, the results of a query can be easily transferred to other VO applications, such as, for instance, Topcat.

\bsp	
\label{lastpage}
\end{document}